\documentclass[longauth, final]{aa}  

\usepackage{hyperref}
\usepackage[utf8]{inputenc}
\usepackage{natbib}
\usepackage{graphicx}
\usepackage{color}
\usepackage{txfonts}
\usepackage{tabularx}

\begin{document}

\title{Gaia Data Release 1: The archive visualisation service}

\author{
A.        ~Moitinho                      \inst{\ref{inst:0119}}\relax
\and A.        ~Krone-Martins                 \inst{\ref{inst:0119}}\relax
\and H.        ~Savietto                      \inst{\ref{inst:0469}}\relax
\and M.        ~Barros                        \inst{\ref{inst:0119}}\relax
\and C.        ~Barata                        \inst{\ref{inst:0119}}\relax
\and A.J.      ~Falc\~{a}o                    \inst{\ref{inst:0263}}\relax
\and T.        ~Fernandes                     \inst{\ref{inst:0263}}\relax
\and J.        ~Alves                         \inst{\ref{inst:0145}}\relax
\and A.F.        ~Silva                        \inst{\ref{inst:0119}}\relax
\and M.        ~Gomes                         \inst{\ref{inst:0119}}\relax
\and J.        ~Bakker
\inst{\ref{inst:ESAC}}\relax
\and A.G.A.    ~Brown                         \inst{\ref{inst:0003}}\relax
\and J.        ~Gonz\'{a}lez-N\'{u}\~{n}ez    \inst{\ref{inst:0048},\ref{inst:0299}}\relax
\and G.        ~Gracia-Abril                  \inst{\ref{inst:0067},\ref{inst:0012}}\relax
\and R.        ~Guti\'{e}rrez-S\'{a}nchez     \inst{\ref{inst:0026}}\relax
\and J.        ~Hern\'{a}ndez                 \inst{\ref{inst:0014}}\relax
\and S.        ~Jordan                        \inst{\ref{inst:0007}}\relax
\and X.        ~Luri                          \inst{\ref{inst:0012}}\relax
\and B.        ~Merin              
\inst{\ref{inst:ESAC}}\relax
\and F.        ~Mignard                       \inst{\ref{inst:0017}}\relax
\and A.        ~Mora                          \inst{\ref{inst:0111}}\relax
\and V.        ~Navarro
\inst{\ref{inst:ESAC}}\relax
\and W.        ~O'Mullane                     \inst{\ref{inst:0014}}\relax
\and T.        ~Sagrist\`{a} Sell\'{e}s       \inst{\ref{inst:0007}}\relax
\and J.        ~Salgado                       \inst{\ref{inst:0122}}\relax
\and J.C.      ~Segovia                       \inst{\ref{inst:0048}}\relax
\and E.        ~Utrilla                       \inst{\ref{inst:0111}}\relax
\and F.        ~Arenou                        \inst{\ref{inst:0005}}\relax
\and J.H.J.    ~de Bruijne                    \inst{\ref{inst:0001}}\relax
\and F.        ~Jansen                        \inst{\ref{inst:0011}}\relax
\and M.        ~McCaughrean           \inst{\ref{inst:ESTEC}}\relax
\and K.S.      ~O'Flaherty                    \inst{\ref{inst:0635}}\relax
\and M.B.      ~Taylor                        \inst{\ref{inst:0503}}\relax
\and A.        ~Vallenari                     \inst{\ref{inst:0004}}\relax
}

\institute{
    CENTRA, Universidade de Lisboa, FCUL, Campo Grande, Edif. C8, 1749-016 Lisboa, Portugal\relax                                                                                                           \label{inst:0119}
    \and Fork Research, Rua do Cruzado Osberno, Lt. 1, 9 esq., Lisboa, Portugal\relax                                                                                                       \label{inst:0469}
    \and UNINOVA - CTS, Campus FCT-UNL, Monte da Caparica, 2829-516 Caparica, Portugal\relax                                                                                                        \label{inst:0263}
    \and University of Vienna, Department of Astrophysics, T\"{ u}rkenschanzstra{\ss}e 17, A1180 Vienna, Austria\relax
            \label{inst:0145}
   \and European Space Astronomy Centre (ESA/ESAC), Camino bajo del Castillo, s/n, Urbanizacion Villafranca del Castillo, Villanueva de la Ca\~{n}ada, E-28692 Madrid, Spain\relax                                          \label{inst:ESAC}
   \and Leiden Observatory, Leiden University, Niels Bohrweg 2, 2333 CA Leiden, The Netherlands\relax                                                                                                         \label{inst:0003}
    \and Serco Gesti\'{o}n de Negocios for ESA/ESAC, Camino bajo del Castillo, s/n, Urbanizacion Villafranca del Castillo, Villanueva de la Ca\~{n}ada, E-28692 Madrid, Spain\relax 
                                                                \label{inst:0048}
    \and ETSE Telecomunicaci\'{o}n, Universidade de Vigo, Campus Lagoas-Marcosende, 36310 Vigo, Galicia, Spain\relax                                                                          \label{inst:0299}
    \and Gaia DPAC Project Office, ESAC, Camino bajo del Castillo, s/n, Urbanizacion Villafranca del Castillo, Villanueva de la Ca\~{n}ada, E-28692 Madrid, Spain\relax                                          \label{inst:0067}
    \and Institut de Ci\`{e}ncies del Cosmos, Universitat  de  Barcelona  (IEEC-UB), Mart\'{i}  Franqu\`{e}s  1, E-08028 Barcelona, Spain\relax                                                                  \label{inst:0012}
    \and Telespazio Vega UK Ltd for ESA/ESAC, Camino bajo del Castillo, s/n, Urbanizacion Villafranca del Castillo, Villanueva de la Ca\~{n}ada, E-28692 Madrid, Spain\relax                                     \label{inst:0026}
    \and European Space Astronomy Centre (ESA/ESAC), Camino bajo del Castillo, s/n, Urbanizacion Villafranca del Castillo, Villanueva de la Ca\~{n}ada, E-28692 Madrid, Spain\relax                              \label{inst:0014}
    \and Astronomisches Rechen-Institut, Zentrum f\"{ u}r Astronomie der Universit\"{ a}t Heidelberg, M\"{ o}nchhofstr. 12-14, D-69120 Heidelberg, Germany\relax                                                 \label{inst:0007}
     \and Laboratoire Lagrange, Universit\'{e} Nice Sophia-Antipolis, Observatoire de la C\^{o}te d'Azur, CNRS, CS 34229, F-06304 Nice Cedex, France\relax                                                        \label{inst:0017}
    \and Aurora Technology for ESA/ESAC, Camino bajo del Castillo, s/n, Urbanizacion Villafranca del Castillo, Villanueva de la Ca\~{n}ada, E-28692 Madrid, Spain\relax                                          \label{inst:0111}
    \and Isdefe for ESA/ESAC, Camino bajo del Castillo, s/n, Urbanizacion Villafranca del Castillo, Villanueva de la Ca\~{n}ada, E-28692 Madrid, Spain\relax                                                     \label{inst:0122}
    \and GEPI, Observatoire de Paris, PSL Research University, CNRS, Univ. Paris Diderot, Sorbonne Paris Cit{\'e}, 5 Place Jules Janssen, 92190 Meudon, France\relax                                             \label{inst:0005}
\and    Scientific Support Office, Directorate of Science, European Space Research and Technology Centre (ESA/ESTEC), Keplerlaan 1, 2201AZ, Noordwijk, The Netherlands\relax                                    \label{inst:0001}
    \and Mission Operations Division, Operations Department, Directorate of Science, European Space Research and Technology Centre (ESA/ESTEC), Keplerlaan 1, 2201 AZ, Noordwijk, The Netherlands\relax              \label{inst:0011}
    \and European Space Research and Technology Centre (ESA/ESTEC), Keplerlaan 1, 2201 AZ, Noordwijk, The Netherlands\relax          
    \label{inst:ESTEC}
    \and EJR-Quartz BV for ESA/ESTEC, Keplerlaan 1, 2201AZ, Noordwijk, The Netherlands\relax                                                                                                                     \label{inst:0635}
    \and H H Wills Physics Laboratory, University of Bristol, Tyndall Avenue, Bristol BS8 1TL, United Kingdom\relax                                                                                              \label{inst:0503}
\and INAF - Osservatorio astronomico di Padova, Vicolo Osservatorio 5, 35122 Padova, Italy\relax                                                                                                             \label{inst:0004}
    }

\date{Received date /
Accepted date}

\abstract{
    The first Gaia data release (DR1) delivered a catalogue of astrometry and photometry for over a billion astronomical sources. Within the panoply of methods used for data exploration, visualisation is often the starting point and even the guiding reference for scientific thought. However, this is a volume of data that cannot be efficiently explored using traditional tools, techniques, and habits. 
}{
    We aim to provide a global visual exploration service for the Gaia archive, something that is not possible out of the box for most people. The service has two main goals. The first is to provide a software platform for interactive visual exploration of the archive contents, using common personal computers and mobile devices available to most users. The second aim is to produce intelligible and appealing visual representations of the enormous information content of the archive.
}{
    The interactive exploration service follows a client-server design. The server runs close to the data, at the archive, and is responsible for hiding as far as possible the complexity and volume of the Gaia data from the client. This is achieved by serving visual detail on demand. Levels of detail are pre-computed using data aggregation and subsampling techniques. For DR1, the client is a web application that provides an interactive multi-panel visualisation workspace as well as a graphical user interface. 
}{
    The Gaia archive Visualisation Service offers a web-based  multi-panel interactive visualisation  desktop in a browser tab. It currently provides highly configurable  1D histograms and 2D scatter plots of Gaia DR1 and the Tycho-Gaia Astrometric Solution (TGAS) with linked views. An innovative feature is the creation of ADQL queries from visually defined regions in plots. These visual queries are ready for use in the Gaia Archive Search/data retrieval service. In addition, regions around user-selected objects can be further examined with automatically generated SIMBAD searches. Integration of the Aladin Lite and JS9 applications add support to the visualisation of HiPS and FITS maps. The production of the all-sky source density map that became the iconic image of Gaia DR1 is described in detail.
}{
    On the day of DR1, over seven thousand users accessed the Gaia Archive visualisation portal. The system, running on a single machine, proved robust and did not fail while enabling thousands of users to visualise and explore the over one billion sources in DR1. There are still several limitations, most noticeably that users may only choose from a list of pre-computed visualisations. Thus, other visualisation applications that can complement the archive service are examined. Finally, development plans for Data Release 2 are presented.
}

\keywords{Galaxy: general -- Astronomical data bases -- Surveys -- Methods: data analysis}

\maketitle

\section{Introduction}

Visual data exploration plays a central role in the scientific discovery process; it is invaluable for the understanding and interpretation of data and results. From analysis to physical interpretation, most research tasks rely on or even require some kind of visual representation of data and concepts, either interactive or static, to be created, explored, and discussed. This is certainly the case of the ESA Gaia space mission \citep{2016A&A...595A...1G},  with its current and planned data releases \citep[e.g.][]{2016A&A...595A...2G}. The particularity of Gaia is the volume -- the number of sources and of attributes per source -- of its data products, which makes interactive visualisation a non-trivial endeavour.

The Gaia Data Releases comprise more than $10^9$ individual astronomical objects, each with tens of associated parameters in the earlier data releases, up to thousands in the final data release, considering the spectrophotometric and spectroscopic data which are produced per object and per epoch. The extraction of knowledge from such large and complex data volumes  is highly challenging. This is a tendency that shows no sign of slowing down in the dawn of the sky surveys such as the LSST \citep{2008arXiv0805.2366I} and the ESA Euclid mission \citep{2011arXiv1110.3193L}. As  several authors have pointed out \citep[e.g.][]{2001Sci...293.2037S, unwin:graphics:2006, hey:fourthparadigm:2009, 2017PASP..129b8001B}, new science enabling tools and strategies are necessary to tackle these data sets;  to allow the best science to be extracted from this data deluge, interactive visual exploration must be performed.

One essential issue is the inherent visual clutter that emerges while visualising these data sets. Although there can be millions or billions of individual entities that can be simultaneously represented in a large-scale visualisation, a naive brute-force system that simply displays all such data would not lead to increased knowledge. In fact, such a system would just hinder human understanding, due to the clutter of information that hides structures that may be present in the data \cite[e.g.][]{Peng04clutterreduction}. Thus, strategies  have to be put in place to address the issues of data clutter and  the clutter of the graphical user interface of the visualisation system \citep[e.g.][]{Rosenholtz05featurecongestion}.

Interactivity is also key for data exploration \citep[e.g.][]{Keim:2002:IVV:614285.614508}. 
The ability to quickly move through the data set (e.g. by zooming, panning, rotating) and to change the representations (e.g. by re-mapping parameter dimensions to colours, glyphs, or by changing the visualised parameter spaces) are  indispensable for productive exploration and discovery of structures in the data.

However, interactivity for large data sets is challenging \citep{2012AN....333..505G} and current approaches require high-end hardware and having the data set locally at the computer used for the visualisation \citep[e.g.][]{2013MNRAS.429.2442H}. In the best cases, these systems are bounded by I/O speed \citep[e.g.][for GPU-based visualisation of  large-scale n-body simulations]{2008arXiv0811.2055S}.

Another essential functionality for visual data exploration is the linking of views from multiple interactive panels with different visualisations produced from different dimensions of the same data set, or even of different data sets \citep[e.g.][]{tukey1977,Jern:2007, Tanaka:2014}. The simultaneous  identification of objects or groups of objects in different parameter projections is a powerful tool for multi-dimensional data exploration \citep[][discusses this in a nice historical perspective]{2012AN....333..505G}.
This surely applies to Gaia with its astrometric, photometric, and spectroscopic measurements \citep{2016A&A...595A...4L, 2017A&A...599A..32V, 2011EAS....45..189K} combined with derived astrophysical information such as the orbits of minor planets  \citep{2016P&SS..123...87T}, the parameters of double and multiple stars  \citep{2011AIPC.1346..122P}, the morphology of unresolved galaxies  \citep{2013A&A...556A.102K}, the variable parameters of stars  \citep{2014EAS....67...75E}, and the classifications and parameters of objects \citep{2013A&A...559A..74B}.

Visualisation is not only for exploring the data, but also for communicating results and ideas. 
One of the most remarkable visualisations of our galaxy was created in the middle of the past century at the Lund Observatory\footnote{\url{http://www.astro.lu.se/Resources/Vintergatan/}}. It is a one-by-two-meter representation of the galactic coordinates of 7000 stars, overlaid on a painting of the Milky Way, represented in an Aitoff projection. This visualisation was produced by Knut Lundmark, Martin  Kesk\"ula and Tatjana Kesk\"ula, and for decades was the reference panorama of our Galaxy. 
Another emblematic and scientifically correct visualisation of the Milky Way was produced from the data gathered by the ESA Hipparcos space mission, and published by ESA in 2013\footnote{\url{http://sci.esa.int/hipparcos/52887-the-hipparcos-all-sky-map/}}. This image represents, in  galactic coordinates, the fluxes of $\sim2.5$ million sources from the Tycho-2 catalogues. The Milky Way diffuse light, mostly created by unresolved stars and reflection or emission in the interstellar medium,  is also represented. It was determined from additional data provided by background measurements from the Tycho star mapper on board  the Hipparcos satellite. Minor additions of known structures not observed by Hipparcos (which just like Gaia was optimised to observe point sources) were made by hand. 
Now, the remarkable Gaia Data Release 1 (DR1) will deliver the next generation of Galactic panoramas, and will set our vision of the Milky Way galaxy  probably for decades to come.

This paper introduces the Gaia Archive Visualisation Service, which was designed and developed to  allow interactive visual exploration of very large data sets by many simultaneous users. In particular, the  version  presented here is tailored to the contents of DR1.
 The paper is organised as follows. First, in Sect.~\ref{sec:sysconc} the system concept is presented and the services described. Then, in Sect.~\ref{sec:deploy} a brief overview of the deployment of the platform is given. Later, Sect.~\ref{sec:contents} presents a thorough description of the visual contents offered by the service and how they were created. Then, Sect.~\ref{sec:other} addresses other visualisation tools with some degree of tailoring to Gaia data. Finally, some concluding remarks and planned developments for the near future are given in Sect.~\ref{sec:conclusions}.

\section{System concept}
\label{sec:sysconc}

In addition to the central functionalities discussed above (e.g. interactivity, large data sets, linked views), many other features are required from a modern interactive visual data exploration facility. The Gaia Data Processing and Analysis Consortium (DPAC) issued an open call to the astronomical community  requesting generic use cases for the mission archive \citep{BrownTN026}. Some of these use cases are related to visualisation, and are listed in Appendix A of this paper. These cases formed the basis for driving the requirements of the Gaia Archive Visualisation Service (hereafter GAVS). 

\paragraph{Visual queries}
In addition, GAVS introduces a new concept of how to deal with database queries: {\it visual queries}. Since the introduction of the Sloan Digital Sky Survey SkyServer and CasJobs infrastructures \citep[e.g.][]{2000AJ....120.1579Y, 2016AJ....151...44D}, astronomers wanting to extract data from most modern astronomical surveys  have been facing the need to learn at least the basics of the Structured Query Language, SQL, or more commonly of the Astronomical Data Query Language, ADQL \citep{2008IVOAADQL}, which is the astronomical dialect of SQL.
These are declarative languages used to query the relational databases at the underlying structure of most modern astronomical data sources. Nevertheless, there are a multitude of querying tasks performed in Astronomy that should not require that these languages be mastered, for example  spatial queries of data lying within polygonal regions of n-dimensional visual representation of tables. Accordingly, GAVS introduces to Astronomy a {\it visual query} paradigm; for Gaia it is possible to create ADQL queries directly from visual representations of data without having to write ADQL directly. 
The visual interface creates a query from a visual abstraction that can be used to extract additional information from the database. The visually created query string can be edited and modified, or coupled to more complex queries. It can be shared with other users or added to scientific papers, thus increasing scientific reproducibility.

\subsection{Architecture}

\begin{figure*}[!htbp]
\centering
    \includegraphics[width=0.80\textwidth]{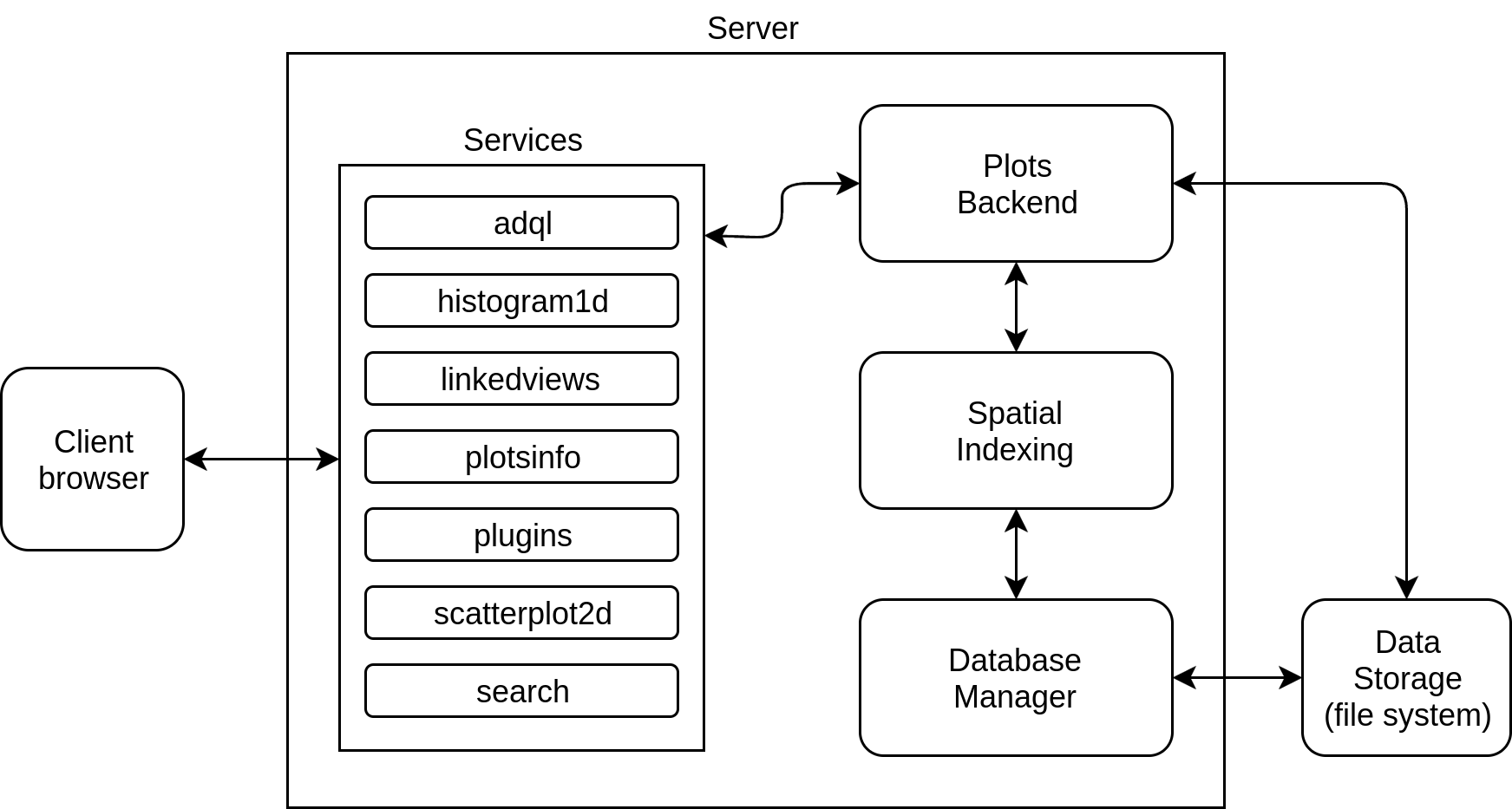}
  \caption{Static architecture diagram for the GAVS Server. It presents the components of the Server (Services, Plots Backend, Spatial Indexing, and Database Manager), how they are connected, and the context within the GAVS.}\label{fig:server_diagram} 
\end{figure*}

From the architectural perspective, one of the fundamental requirements is that the interactive visual explorations of the whole archive should be possible with the common laptops, desktops, and  (if possible) mobile devices available to most users. 

The GAVS described here addresses this architectural issue by adopting a web service pattern. A server residing near the data is responsible for hiding as far as is possible the complexity and volume of the Gaia archive data from the user web interface. This avoids huge brute force data transfers of the archive data to the remote visualisation display that would congest the servers, the network, the user machine, and that in the end would not convey any additional scientific information. In a way,  this reaffirms the concept of `bring the computation to the data' \citep{hey:fourthparadigm:2009}.
However, the server can create an additional pressure on the archive, especially when several users access the service in parallel. To alleviate this pressure, the service includes caching mechanisms to prevent performance penalties from repeated requests and/or re-computations of the same data. This caching mechanism is active whether the request is being performed by the same user or not.
While the server design is not tied to a specific hardware configuration, it pursues a scalable solution that can run on modest hardware (see Sect.~\ref{sec:deploy}).

The server was implemented as a Java EE application, designed to run in Apache Tomcat web containers. This application has two main functions, processing dynamic requests for interactive visualisation purposes and delivering static content to the user’s browser (images, CSS files, and client scripts).

The client side is a web application, designed in Javascript, HTML, and CSS to run in a web browser. Chrome and Firefox are the recommended platforms as these were the platforms used to test the service. Still, the client-side application should be compatible with any modern web browser.

The next two sections detail the server and client components of GAVS.

\subsection{Visualisation Server}
\label{sec:server}


The structure of the Visualisation Server is depicted in Fig. \ref{fig:server_diagram}. It is responsible for receiving and interpreting requests related to the different provided services (see Table \ref{tab:serv_service}) and responding accordingly. 

The  server's components are divided into two different levels: the Services and the Plots Backend. The Services component receives REST requests \citep{Fielding:2000}, and performs checks to ensure the validity of the request and of its parameters. Then, it converts these parameters from the received text format to the correct abstractions and makes the necessary calls to the Plots Backend. Finally, it processes the answers of the Backend and adapts the replies to the visualisation client. The Plots Backend component processes the requests interpreted by the Services at a lower level, and is  responsible for processing data, generating static images and image tiles, and calling further libraries as needed.

Spatial Indexing is a specialised module for indexing data in a spatial way, supporting an arbitrary number of dimensions and  data points. Each specific visualisation will have its own separate index pre-computed (e.g. a scatter plot of galactic longitude and latitude will have an index built from those two coordinates). This pre-computation is key for providing interactivity. Scalability tests with the current implementation of the indexing were performed, indicating the feasibility of treating more than $2\times10^9$ individual database entries using a normal computer on the server side (16GB of RAM with a normal $\sim500$MB/s SSD attached to the SATAIII bus).

 The indexing works in the following manner. First, the minimum and the maximum values are determined for each dimension of the data space being indexed. Based on this information, the root page of a tree is created. Then, data points are inserted into the root page one by one. If the number of data points in a page exceeds a certain configured threshold, the page is divided into children and the data points are also split among the child pages. The division of a page is performed by dividing each dimension by two; therefore, the number of children after a split will be $2^d$. In one dimension, each page is divided into two child pages, in two dimensions each page is divided into four child pages, and so on. When querying the index for data, only the pages that intersect the query range (in terms of area  or volume, depending on the number of the dimensions of the index) are filtered, reducing significantly the amount of processing required.

The ideal threshold for the number of data points per page must be chosen taking into consideration that each page will always be retrieved from the database as a whole block. Accordingly, if this number is too high the amount of data retrieved from the database per request will be too big, even for spatial queries in small regions of the data space. On the other hand, if this threshold is too low, spatial queries will request a very high number of small blocks from the database. Both scenarios can hinder the performance of the application and prevent a satisfactory user experience on the client side. Our tests indicate that limiting the number of data points per page to the range $6-12\times10^4$ yields satisfactory response time for interactive visualisation.

Inside each page the data points are divided among different levels of detail. This is done for two main reasons: first, to prevent data crowding while producing the visual representation of the data (care is taken to avoid cropping or panning issues in the representation), and second, to keep the number of individual data points to be passed to the visualisation client and to be represented on the screen at a limit that permits the client side to experience interactivity.


Levels of details are numbered from 0 to n, with 0 being the level of detail containing the fewest data points or, in our terminology, the lowest level of detail. In our representation, the levels of detail are cumulative, i.e. level $n+1$ includes all the data points of level $n$. Nonetheless, the data points are not repeated in our data structure, and any query processing just accumulates the data of each previously processed level up to the requested level of detail. The number of data points at each level of detail is $2^d$ times the number of the previous one. For example, if the level of detail 0 has 500 data points, the level of detail 1 will have 2,000, the level of detail 2 will have 8,000 and so on. There are several ways in which the selection of points can be performed, but the most direct one, a simple random sampling, is known to present several advantages for visualisation. As discussed by \cite{4376143}, among other features, it keeps spatial information, it can be localised, and it is scalable.


Storing the data structures and providing further querying functionalities require a final component, a Database Manager. The visualisation service described in this paper can use any Database Manager (e.g. MongoDB\footnote{https://www.mongodb.com/}, OrientDB\footnote{http://orientdb.com/}) that can provide at least the two most basic required functions, storing and retrieving data blocks. A data block is a string of bytes with a long integer number for identification. The internal data organisation within the database is irrelevant for indexing purposes. The present implementation of GAVS, tailored to DR1, adopts our own Java-optimised NoSQL Database.

While tree indexation is common in multi-dimensional data retrieval, the specific indexing and data serving schemes  developed here are, to the best of our knowledge, unique in systems for interactive visualisation of (large) astronomical tables.

\begin{table}[ht]
\caption{Role of the services provided by the server side.}
\label{tab:serv_service}
\begin{tabularx}{0.45\textwidth}{XX}
\\\hline
    \noalign{\smallskip}
Service & Role \\\hline\noalign{\smallskip}
adql & ADQL query generation and validation \\    \noalign{\smallskip}\hline    \noalign{\smallskip}

histogram1d & 1D histogram data manipulation and static 1D histogram generation \\    \noalign{\smallskip}\hline    \noalign{\smallskip}

linkedviews & Linked views for point selections and data subsets \\    \noalign{\smallskip}\hline    \noalign{\smallskip}

plotsinfo & Information on plots metadata (dimensions, axes names, axes limits, and more)
 \\    \noalign{\smallskip}\hline    \noalign{\smallskip}

plug-ins & Data from JS9 and Aladin plug-ins \\    \noalign{\smallskip}\hline    \noalign{\smallskip}

scatterplot2d & 2D scatter plot image generation, both dynamic and static \\    \noalign{\smallskip}\hline    \noalign{\smallskip}

search & Name search in external services (CDS/Sesame) \\    \noalign{\smallskip}\hline    \noalign{\smallskip}
\end{tabularx}
\end{table}

\subsection{Web client}

\begin{figure*}[!htbp]
\centering
    \includegraphics[width=0.80\textwidth]{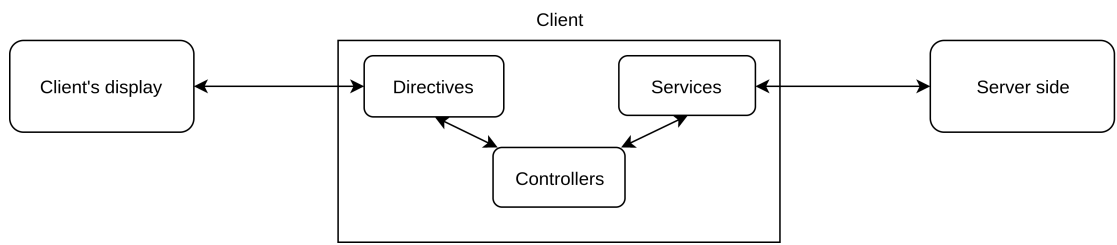}
  \caption{Client architecture diagram providing the structure and context in GAVS. The Client is structured in a Model-View-Controller
design expressed in the Directives, Controllers, and Services modules.}\label{fig:cli_diagram}
\end{figure*}

\begin{table}[ht]
\caption{Role of the services, directives, and controllers that compose the architecture of the client side.}
\label{tab:cli_service}
\begin{tabularx}{0.45\textwidth}{XX}
\\\hline
    \noalign{\smallskip}
Service & Role \\\hline\noalign{\smallskip}
adql & ADQL query requests\\    
    \noalign{\smallskip}

histogram1d & 1D histogram requests\\    
    \noalign{\smallskip}

createVisualizations & Creation of visualisations by user file requests\\    
    \noalign{\smallskip}

plug-ins & JS9 and Aladin lite plug-ins requests\\    
    \noalign{\smallskip}

scatterplot2d & 2D scatter plot requests\\    
    \noalign{\smallskip}

visualizationsService & Created visualisations request\\    
    \noalign{\smallskip}\hline    \noalign{\smallskip}
Directive & Role\\\hline\noalign{\smallskip}
I/O modals & Allows  user to communicate with the system\\    
    \noalign{\smallskip}

aladin & Shows an Aladin lite window\\    
    \noalign{\smallskip}

js9 & Shows a JS9 window\\    
    \noalign{\smallskip}

main & Creates the main page and the gridster windows\\    
    \noalign{\smallskip}

plotWithAxes & Creates an abstract plot that can assume any available type: Histogram 1D or Scatterplot 2D\\    
    \noalign{\smallskip}\hline    \noalign{\smallskip}
Controller & Role\\
    \hline\noalign{\smallskip}
adqlController & Controls the adql I/O\\    
    \noalign{\smallskip}

aladinLiteController & Controls the Aladin lite window\\    
    \noalign{\smallskip}

modalsControllers & Controls the I/O modals\\    
    \noalign{\smallskip}

histogram1DController & Controls the histogram 1D windows\\    
    \noalign{\smallskip}

js9Controller & Controls the js9 window\\
    \noalign{\smallskip}

mainController & Controls the main window functions and the gridster windows\\    
    \noalign{\smallskip}
 
stateController & Controls the save and restore state\\
    \noalign{\smallskip}
 
scatterPlot2DController & Controls the scatterplot 2D windows\\    
    \noalign{\smallskip}\hline    
 
 \end{tabularx}
\end{table}

The structure of the web client is depicted in Fig. \ref{fig:cli_diagram}. The client is responsible for the user interaction with the visualisation service and thus for the communication between the user's computer and the visualisation server. 

The Client is a single-page application structured in a Model-View-Controller (MVC) design pattern. Accordingly, the Client is divided into three major components:

\begin{itemize}
\item the directives that manipulate the HTML and thus serve data to the client’s display;
\item the services that communicate directly with the server side through REST requests;
\item the controller that works as the broker between the services and the directives.
\end{itemize}

The components, available services, and the specifications of each individual role are described in Table~\ref{tab:cli_service}.

Grids of windows providing different functionalities can be created on the client web page using the gridster.js\footnote{\url{http://dsmorse.github.io/gridster.js/}} framework. Using these windows, the Visualisation Service deployed for DR1 provides the following types of plots: 1D histograms, 2D scatter plots, and the JS9\footnote{\url{http://js9.si.edu/}} (FITS viewer) and Aladin lite\footnote{\url{http://aladin.u-strasbg.fr/AladinLite/}} (HiPS  viewer) plug-ins. Options and configurations for each plot are available in modal windows that appear superposed on the main web page when requested.


The 1D histograms are visualised (but not computed) using the d3.js\footnote{\url{https://d3js.org/}} library. This library provides tools for drawing the histogram bins and axes. For 1D histograms, the client requests the bin values to the Server, specifying the number of bins and the maximum and minimum limits over which to compute the histogram. The Server then calculates the number of points in each bin and replies these values to the client using a JSON object. Performance at the server side is improved by not counting every single data point in the data set. If the limits of a data page in the index fall within the limits of a bin, the pre-computed total number of points of the page is used, instead of iterating over every data point. This provides quick response times, allowing to interactively change bin limits and sizes,  even for the over one billion points in DR1. 

The 2D scatter plots are produced using the leaflet\footnote{\url{http://leafletjs.com/}} interactive map library. This library is specialised in tile-based maps and has a small code footprint. The Server application generates the tiles from projections of the data based on client-side requests. The client side then uses these tiles via leaflet to display them to the user. The axes of the scatter plots are created following the same underlying logic and libraries as the 1D histograms, providing a homogeneous user experience throughout the visualisation service. Finally, the client-side application also supports additional overlays with interactive layers and vector objects.

\section{Deployment}
\label{sec:deploy}

The entire development and prototyping phase of the service was performed using a virtual machine infrastructure at ESAC, together with a physical set-up at the Universidade de Lisboa. 


The visualisation web service is deployed at ESAC in a dedicated physical machine. This service came online together with the Gaia DR1. It has been in continuous operation since the moment the archive went public on September 14, 2016.

The service is accessible through the Gaia Archive portal\footnote{\url{http://gea.esac.esa.int/archive/}}, in a special pane dedicated to the online visual exploration of the data release contents. It can also be accessed via a direct link\footnote{\url{http://gea.esac.esa.int/visualisation}}.

 
The fundamental configuration and characteristics of the operational infrastructure are

\begin{itemize}
\item CPU: Intel(R) Xeon(R) E5-2670 v3 @ 2.30GHz, 16 cores;
\item Memory: 64 gigabytes;
\item Storage: 3 TB SSD;
\item Application server: Apache Tomcat 8;
\item Java version: 1.8.
\end{itemize}

The software and hardware deployed for the Visualisation Service proved robust. It has not crashed even once in the several months it has been online, despite several heavy access epochs, and also considering that the service is sustained by a single physical machine.

 In the first four hours after starting online operations, the visualisation service had already served more than 4286 single users. These users created and interacted with 145 1D histograms and 5650 2D scatter plots,   which triggered the generation of $>1.5\times10^6$ different tiles\footnote{The caching mechanism prevents a tile from being created twice.}.



By the end of the DR1 release day, over 7500 individual users had been logged and interacted with the visualisation service.

\section{Contents produced for DR1}
\label{sec:contents}

In the service deployed for DR1, the visualisation index pre-computations (sect.~\ref{sec:server}) are determined by the GAVS operator. Hence, the GAVS portal serves a predefined list of scatter plots and histograms:
\begin{description}
    \item 1D histograms
    \begin{itemize}
        \item GDR1 data: galactic latitude; galactic longitude; G mean magnitude; G mean flux
        \item TGAS data: parallax; proper motion in right ascension; proper motion in declination; parallax error; proper motion modulus
    \end{itemize}
    \item 2D scatter plots
    \begin{itemize}
        \item GDR1 data: galactic coordinates; equatorial coordinates; ecliptic latitude and longitude
        \item TGAS data: parallax error vs. parallax; proper motion in declination vs. proper motion in right ascension; colour magnitude diagram (G-Ks vs. G, with Ks from 2MASS)
    \end{itemize}
  \end{description}

\begin{figure*}[!htbp]
\centering
    \includegraphics[width=0.80\textwidth]{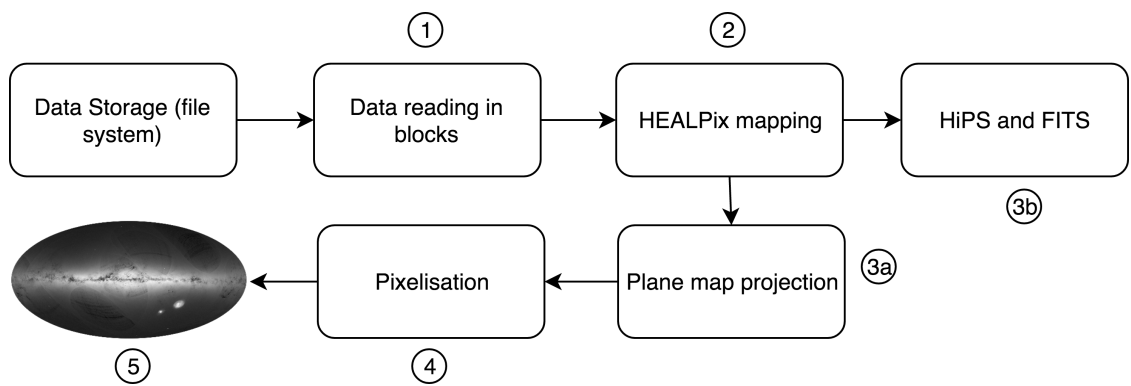}
  \caption{Outline of the steps followed when  creating visualisations for distribution and for viewing with external applications. It covers the creation of the DR1 poster image in various formats as well as HiPS and and FITS files.}\label{fig:map_pipe}
\end{figure*}

In addition to the interactive scatter plots and histograms that can be explored in the visualisation portal, the service has also produced content  for distribution or for viewing with other specialised software. It is made available at GAVS under the `Gallery' menu item. Here we list that other content and briefly describe how it was created:
\begin{itemize}
\item All-sky source density map in a plane projection. This is the DR1 poster image shown in \cite{2016A&A...595A...2G} and available in several sizes at \url{http://sci.esa.int/gaia/58209-gaia-s-first-sky-map/}, also with annotations;
\item A similar map, but for integrated (logarithmic) G-band flux. It is shown and discussed below;
\item Several zoom-ins of regions of interest, re-projected centred on those regions. Employed as presentation material. Two examples are shown and discussed below;
\item All-sky HiPS maps of source density and integrated (logarithmic) G-band flux for viewing with the Aladin Lite plug-in  at the Archive Visualisation Service;
\item All-sky  low-resolution FITS map in a cartesian projection with WCS header keywords for viewing with the JS9 plug-in in the Archive Visualisation Service;
\item FITS images of selected regions in orthographic projection with WCS header keywords;
\item Large format all-sky source density and integrated flux maps for projection in planetaria.
\end{itemize}

Images are produced with the pipeline presented in Fig.~\ref{fig:map_pipe}. The pipeline is written in python.
The input data are stored in  tabular form in the file system. 
Schematically, the steps are as follows:
\begin{enumerate}
    \item Data are read in blocks. This allows  images to be produced from tables of arbitrary sizes, larger than would fit in memory, as long as there is enough disc storage space. The pipeline uses the python package Pandas\footnote{\url{http://pandas.pydata.org}}  in this process. 
    \item\label{item:hpx_map} The computation of the statistic to be visualised requires partitioning the celestial sphere in cells. The Healpy\footnote{\url{https://github.com/healpy/healpy}} implementation of the Hierarchical Equal Area isoLatitude Pixelation \citep[HEALPIX, ][]{gorski2005healpix} tessellation is used for this purpose. Each source is assigned a HEALPix from its sky coordinates.
    The statistic can be simply the number of sources in the cell, the integrated luminous flux of sources in the cell, or any other quantity that can be derived from the source attributes listed in the input table. The statistics determined for the data blocks are added to a list of values for each HEALPix. Averaged or normalised statistics (e.g. number of sources per unit area) are only computed in the end to avoid round-off errors. Finally, the central sky coordinates of each HEALPix are determined and a list of the statistics for those coordinates is produced.
    
    \item[3a.] The statistics determined on the sphere are represented on a plane. Because the sphere cannot be represented on a plane without distortion, many approaches exist for map projections \citep{Synder1993}. The Hammer projection, used to produce the DR1 image, is known to be an equal-area projection that reduces distortions towards the edges of the map. The projection results in x,y positions in a 2:1 ratio with x confined to (-1,1). The zoomed images use an orthographic--azimuthal projection, which is a projection of points onto the tangent plane.

    \setcounter{enumi}{3}
    \item The x,y coordinates of the map projection are re-sampled (scaled and discretised) onto a 2D matrix with a specified range (image dimensions in arcminutes) and number of pixels in each dimension. The number stored in each pixel corresponds to the combined statistics of the HEALpix that fall in the pixel. Because the HEALPix and pixels have different geometries, they will not match perfectly. It is thus important that the pixel area should be substantially larger than the HEALPix area, i.e. that each pixel includes many HEALPix to minimise artefacts due to differences in the areas covered by both surface decompositions.  In the case of the Hammer projection, we have found that a pixel area 32 times larger that the HEALPix area will keep artefacts at the percent level. Given the 2:1 aspect ratio of the Hammer projection, this corresponds to an average of 8$\times$4 HEALpix per pixel.

    \item The image is rendered from the matrix built in the previous step. There are many libraries available, but only a few produce images with 16 bit colour maps. This is required to produce high-quality images with enough colour levels (65536 levels of grey, compared to 256 levels for 8 bit images) to go through any post-processing that might be desirable for presentation purposes. Here the pyPNG\footnote{\url{https://pythonhosted.org/pypng/}} package is used. The output is a PNG image.

    \item[3b.] HiPS and FITS image files are produced. Healpy can create FITS files with HEALPix support (embedded HEALPix list and specific header keywords) directly from the HEALPix matrix produced in step~\ref{item:hpx_map} of the image pipeline. HiPS images were then created from the HEALPix fits file with the Aladin/Hipsgen code following the instructions in  \url{http://aladin.u-strasbg.fr/hips/HipsIn10Steps.gml}. The input data were mapped into HiPS tiles, without resampling, using the command java -Xmx2000m -jar Aladin.jar in=`HealpixMap.fits'.  While this method allows a quick and easy creation of HIPS files, it does not handle well very high resolutions. To illustrate the issue, for a nside of $2^{13}$=8192 the HEALPixs array has a length of 805306368, while for a nside of  $2^{14}$ it has a length of 3221225472. The DR1 HiPS maps provided in the Visualisation portal have a base nside of 8192.
    Regarding JS9, a javascript version of the popular DS9 FITS viewer\footnote{\url{http://ds9.si.edu}}, the current version does not support HEALPix FITS files. For JS9, the HEALPIx map was directly projected on the cartesian plane and converted to FITS using the Astropy\footnote{\url{http://www.astropy.org/}} FITS module astropy.io.fits.
\end{enumerate}

\begin{figure*}[!htbp]
\centering
    \includegraphics[width=0.8\textwidth]{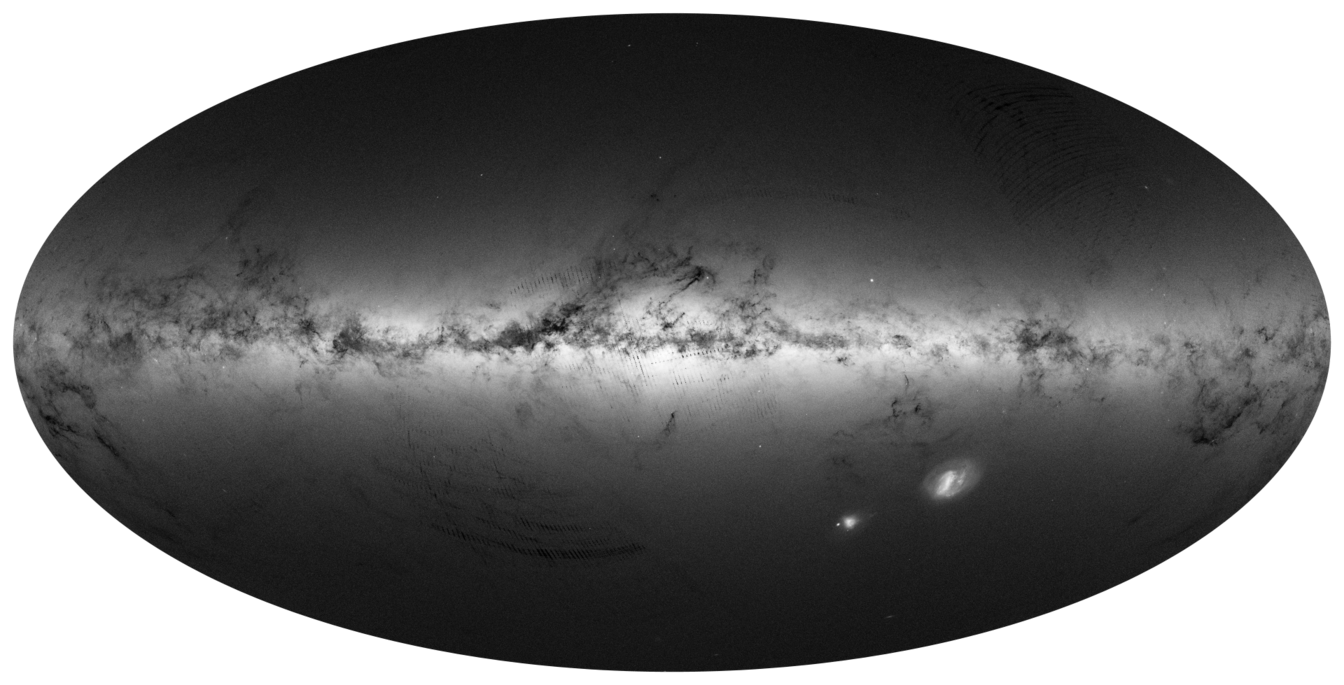}\\
    \includegraphics[width=0.8\textwidth]{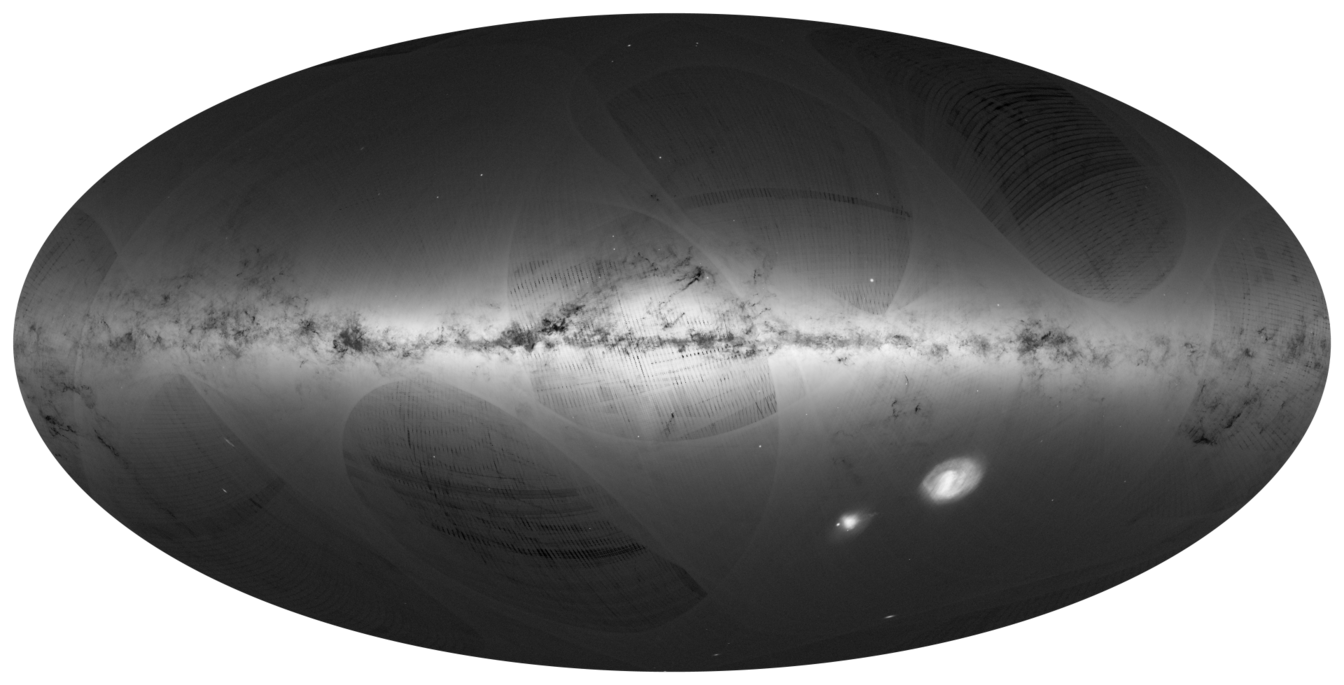}
    \includegraphics[width=0.8\textwidth]{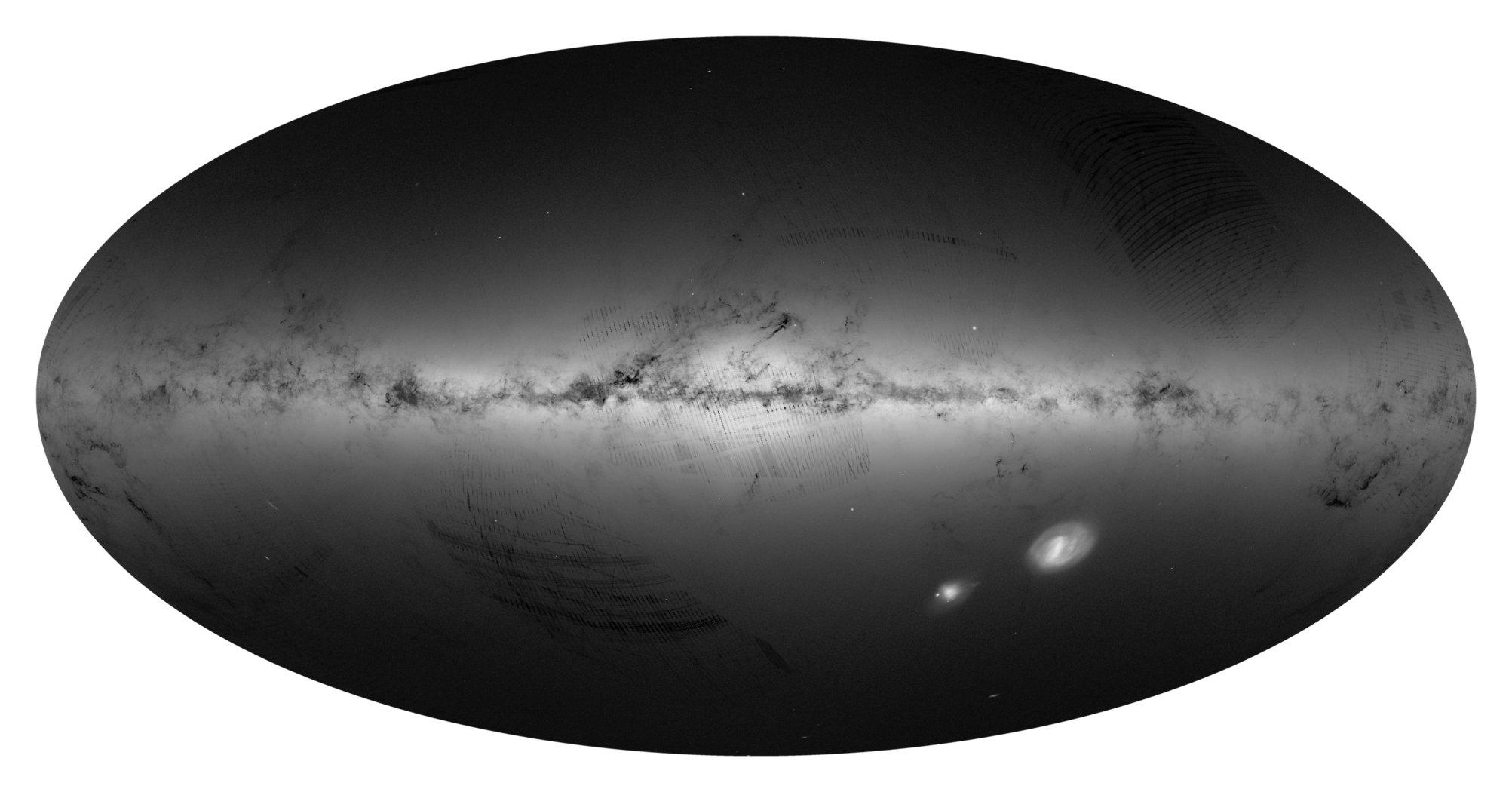}
  \caption{All-sky maps of DR1: Integrated flux (top), density (middle), density for sources brighter than G=20 mag (bottom)}\label{fig:allsky}
\end{figure*}

The DR1 poster image \citep{2016A&A...595A...2G} is available in several sizes\footnote{\url{http://sci.esa.int/gaia/58209-gaia-s-first-sky-map/}}, also with annotations. It 
is a Hammer projection of the Galactic plane represented in galactic coordinates. This specific projection was chosen in order to have the same area per pixel. 

The images of different sizes are scaled versions of a baseline image of  8000$\times$4000 pixels, which corresponds to an area of$\sim 5.901283423$ arcmin$^2$ per pixel. As explained above, the plane projected images are created from higher resolution HEALPix matrices. In this case, an NSIDE = 8192 was used, which corresponds to an area of 0.184415106972 arcmin$^2$ per HEALPix or ($\sim$ 8$\times$4) 32 HEALPix per pixel.

The greyscale represents the number of sources/arcmin$^2$. In \cite{2016A&A...595A...2G} a scale bar is presented together with the map. The maps mentioned above, which are available at the ESA website, are based on a logarithmic scale followed by some post-processing fine-tuning of the scale with an image editing program.
As noted in \cite{2016A&A...595A...2G}, the scales were adjusted in order to highlight  the rich detail of Galactic plane and the signature of the Gaia scanning.

The maximum density is slightly higher than 260 sources/ arcmin$^2$ 1.000.000 sources/degr$^2$ 
The minimum is 0, but this mostly due to gaps in certain crowded regions where no sources have been included in DR1,  noticeably the stripes close to the Galactic centre. Not considering these missing parts with zero density, the minimum at this resolution is about $\sim$300 sources/degr$^2$ .

As mentioned above, an all-sky logarithmic integrated G flux map was also produced. It is shown in Fig.~\ref{fig:allsky} together with the density map for comparison. 
While the density map highlights dense groups of stars, even  very faint stars at the limiting magnitude, the flux map can highlight sparse groups of bright stars.
This explains why the density map is so full of detail. Many very faint but dense star clusters and nearby galaxies are easily seen. Features in the dust distribution also become prominent as they create pronounced apparent underdensities of stars. 
It also explains why the striking Gaia scanning patterns in the density map are mostly absent in the flux-based map. As discussed  in \cite{2016A&A...595A...2G}, the patterns are an effect of incompleteness which affects mostly the faint end of the survey. This is confirmed with the density map in the lower panel of Fig.~\ref{fig:allsky}, which was built from sources brighter than G=20 mag and shows many fewer scanning footprints. 

\begin{figure*}[!htbp]
\centering
    \includegraphics[width=0.45\textwidth]{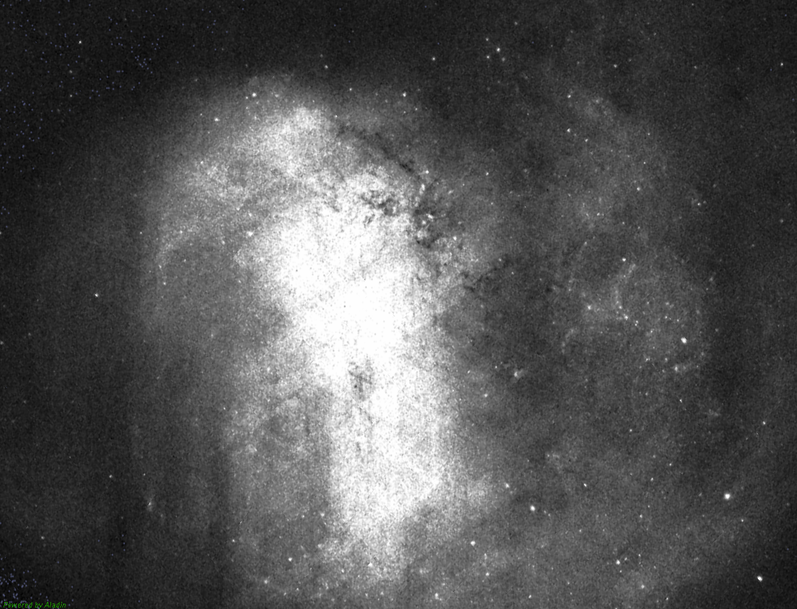}
    \includegraphics[width=0.45\textwidth]{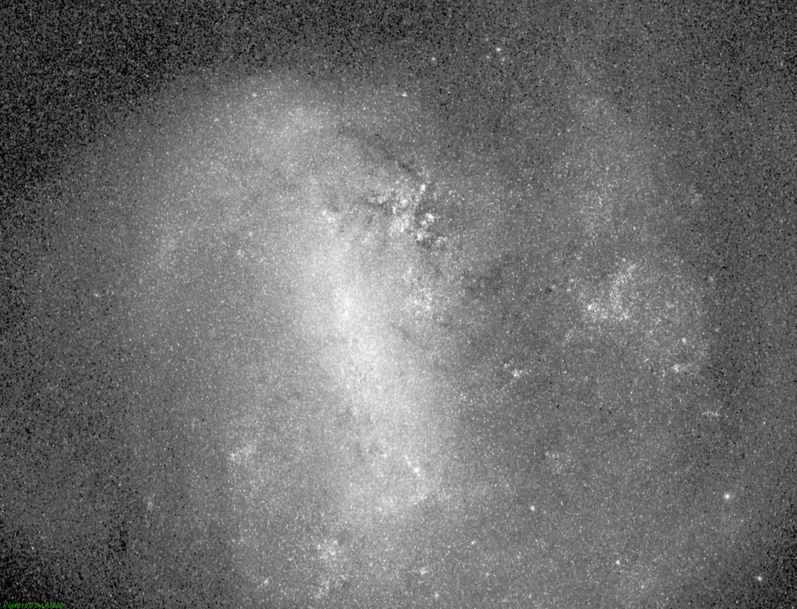}
  \caption{$12\degr\times 10\degr$ density (left) and integrated flux (right) maps of the LMC.}\label{fig:lmc}
\end{figure*}

This illustrates how these density and flux-based maps provide complementary views, where one reveals structures that are not seen in the other. This is further illustrated in Fig.~\ref{fig:lmc}, which is a zoom into a field of $~\sim 12\degr\times 10\degr$ centred on the LMC. Here the LMC bar and arms are seen differently in the two images. The density map displays scanning artefacts, specially in the bar, but also reveals many faint star clusters and clearly delineates the extent of the arms.  The structure of the bar and the 30 Doradus region (above the centre of the images) are better revealed by the bright stars that dominate the flux-based image. It is worth noting that despite its photo-realism, this is not a photograph, but a visualisation of specific aspects of the contents of the DR1 catalogue.

\begin{figure*}[!htbp]
\centering
    \includegraphics[width=0.45\textwidth]{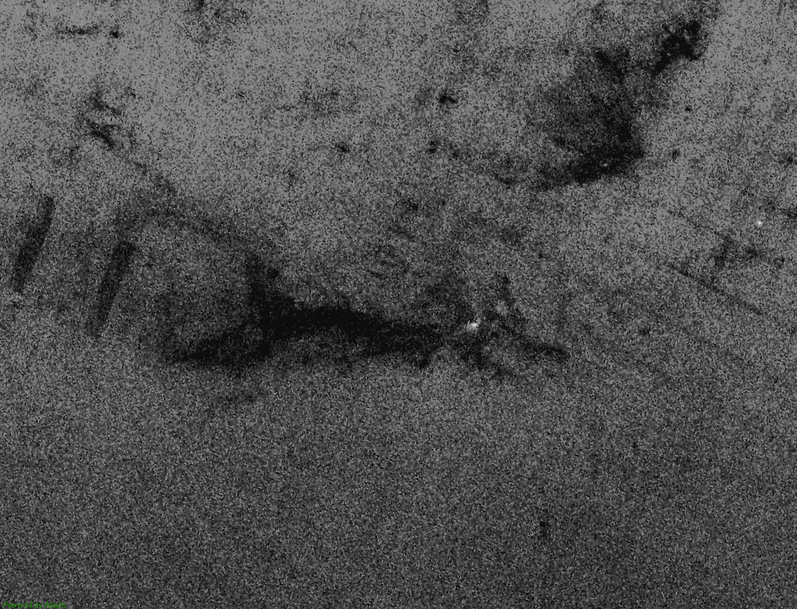}
    \includegraphics[width=0.45\textwidth]{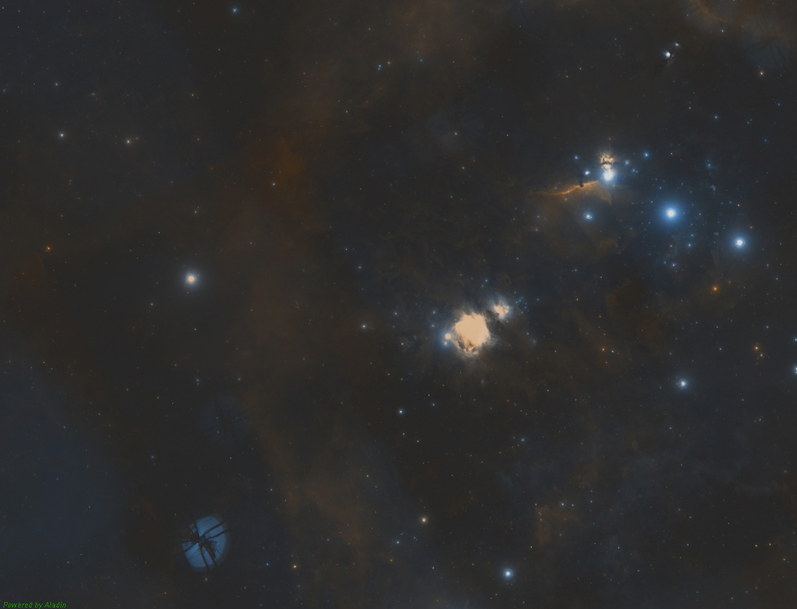}\\
    \includegraphics[width=0.45\textwidth]{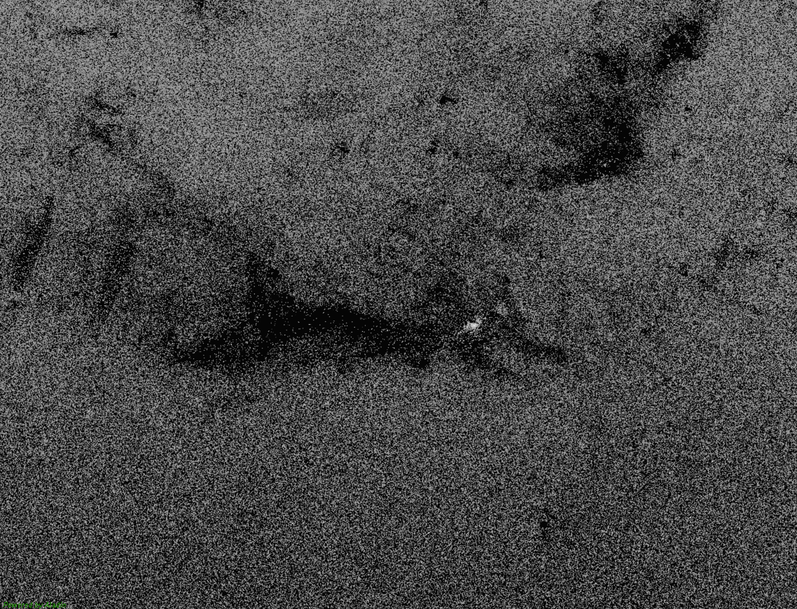}
    \includegraphics[width=0.45\textwidth]{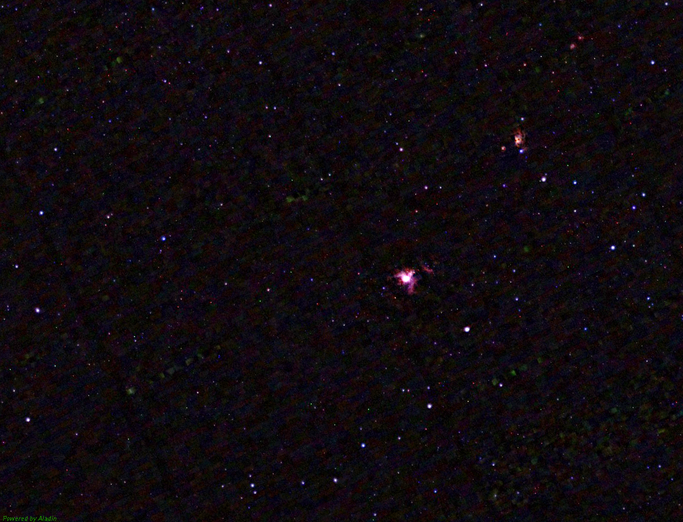}
  \caption{$15\degr\times 11\degr$ field centred on the Orion-A region. Density (top left) and integrated flux (bottom left) maps with DR1 data. Coloured DSS (top right) and 2MASS (bottom right) images of the same field. The DR1 images reveal a cat-like structure created by an extinction patch, highlighting how the mere positions in DR1 can already reveal structures not seen in currently available optical and near-infrared images.}\label{fig:orion}
\end{figure*}

While the astrophysical interest of Orion cannot be overstated, it also provides a highlight of how even though Gaia is an optical mission, the mere positions of the stars published in DR1 can reveal structures not yet revealed by other surveys. Figure~\ref{fig:orion} shows four panels of a $\sim 15\degr\times 11\degr$ field centred on the Orion-A cloud. The two panels on the left are DR1 density (upper) and flux (lower) maps. The panels on the right are coloured  DSS (upper) and 2MASS (lower) images of the same field. The sources in the DR1 maps delineate a distinctive extinction patch that closely resembles a cat\footnote{At public presentations, some members of the audience have suggested that it is a fox. We are currently re-analysing the data and taking a deeper look into this issue.} flying or jumping stretched from the left to the right, with both paws to the front. This structure is not seen in currently available optical and near infrared images, except for the shiny `nose' and the cat's `left eye'.


Finally, to end this section, large 16384$\times$8192 pixel density and integrated flux maps in a cartesian projection have been produced for display in planetaria. They are currently employed in several Digistar\footnote{\url{https://www.es.com/digistar/}} planetaria around the world.

\section{ Workflows}
\label{sec:workflows}

An online GAVS Quick Guide can be consulted under the `Help' menu button. The guide describes the full set of functionalities offered by GAVS. It includes explanations of the basic capabilities such as adding new visualisation panels, types of visualisations, presets, and configurations. It also covers more advanced features such as creating (and sharing) geometrical shapes for marking regions of interest, overlaying catalogues of objects, and generating ADQL visual queries.

This section gives a few  examples of workflows using GAVS. More examples and details on the user interface can be found  in the online guide.

\begin{figure}[!htbp]
\centering
    \includegraphics[width=0.45\textwidth]{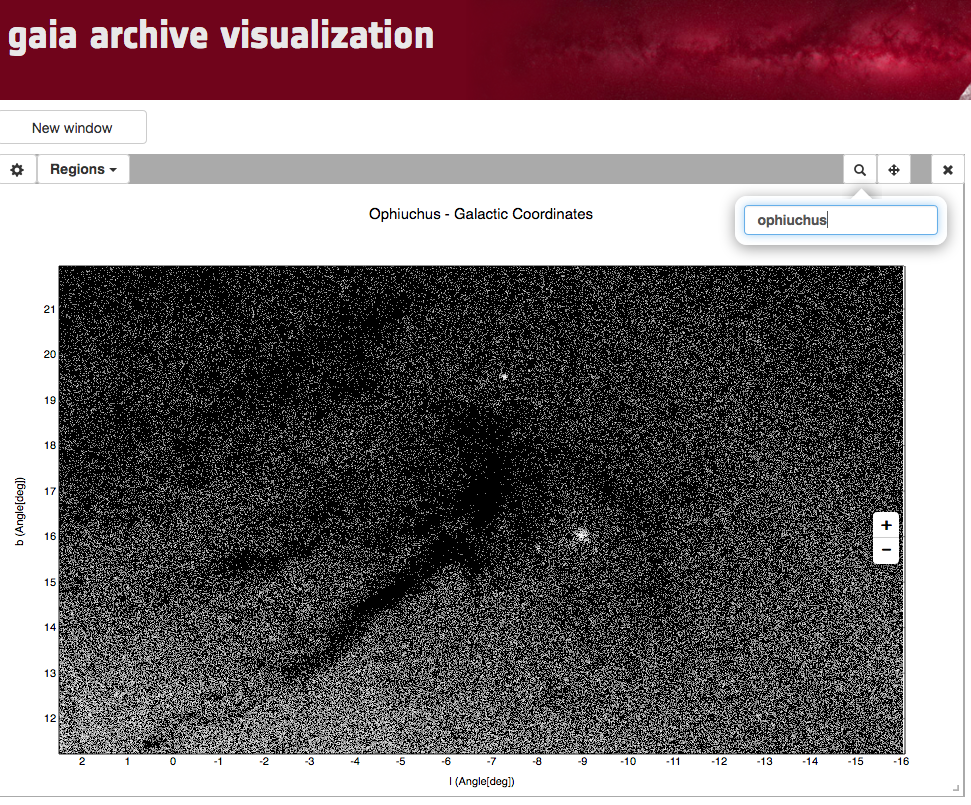}
    \includegraphics[width=0.45\textwidth]{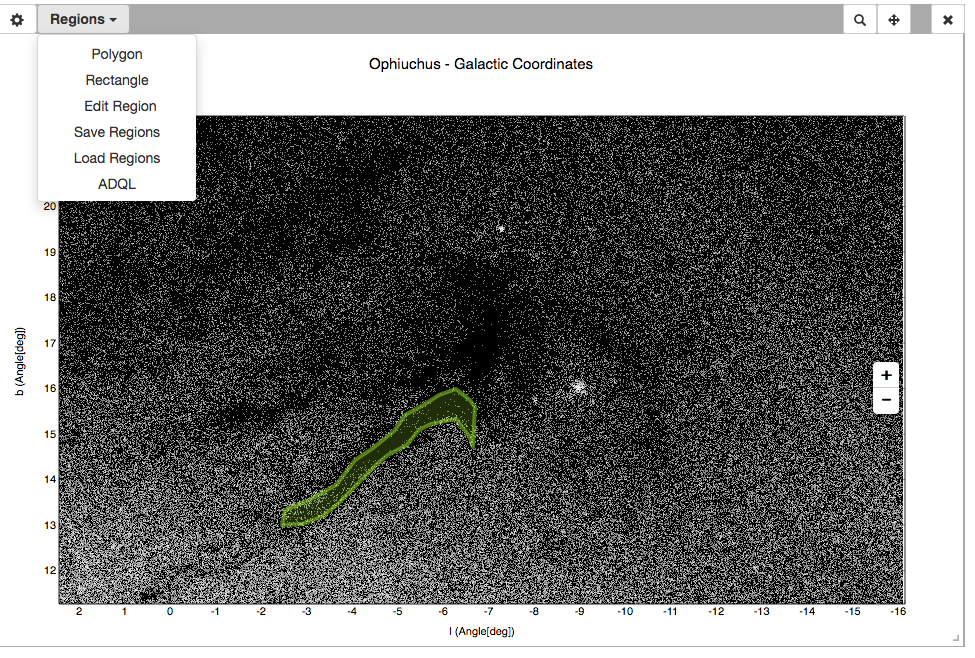}\\
    \includegraphics[width=0.45\textwidth]{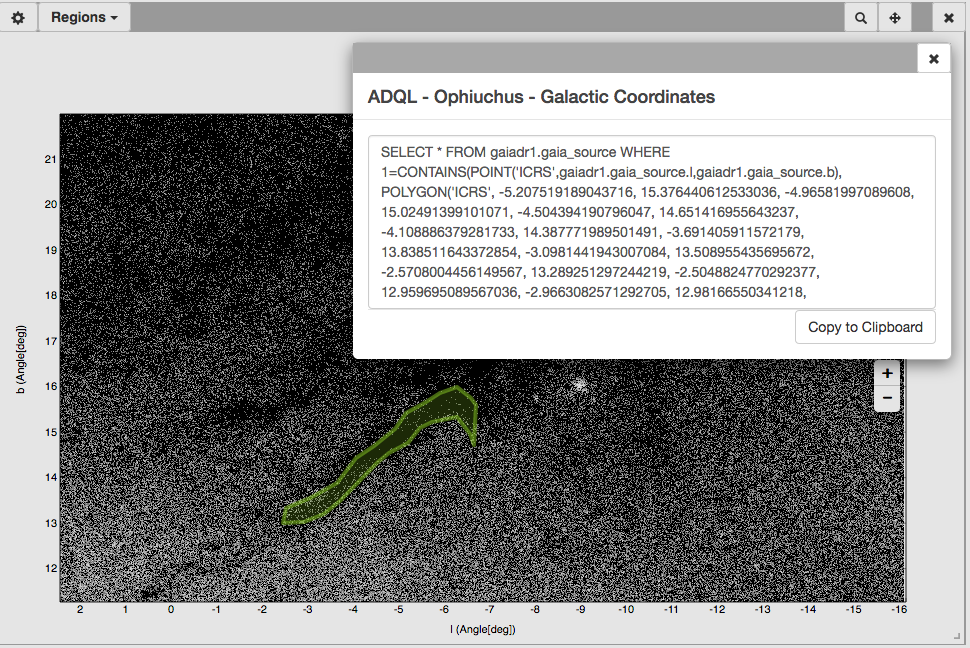}
  \caption{Workflow for producing a  visual ADQL query. Top panel: The user selects a region to centre the view (top). Middle panel: The selected polygonal region is shown in green. Region menu with the ADQL functionality also displayed. Bottom panel: The resulting ADQL query.}\label{fig:adql_oph}
\end{figure}


Figure~\ref{fig:adql_oph} illustrates a workflow for centring on a field of interest (using the Sesame name resolver), marking a region, and producing an ADQL query that can be pasted into the archive query interface. In the first step (top panel) the user clicks on the lens icon in the top right corner of the window and enters the name of a region or object (in this example, Ophiuchus). The visualisation window will centre the field on the region if the CDS Sesame service can resolve the name. Alternatively, instead of a name, central coordinates can be used as input. Afterwards, the user clicks on the `Regions' menu of the visualisation window  in the top left, and selects a polygonal region or rectangular region. The user then creates the region using the mouse,  e.g. by clicking on each vertex of the polygon, and closing the polygon at the end (middle panel). Finally, the user clicks again on the `Regions' menu and selects the `ADQL' option. This will result in the creation of an ADQL query that is presented to the user (bottom panel). Behind the scenes, the software validates the resulting ADQL query before  presenting it to the user, assuring that it is correctly constructed. The user can now run this query as is at the Gaia Archive search facility, or can customise it (e.g. to select which data columns to retrieve from the table or to perform a table join) before submitting it to the archive.

\begin{figure}[!htbp]
\centering
    \includegraphics[width=0.45\textwidth]{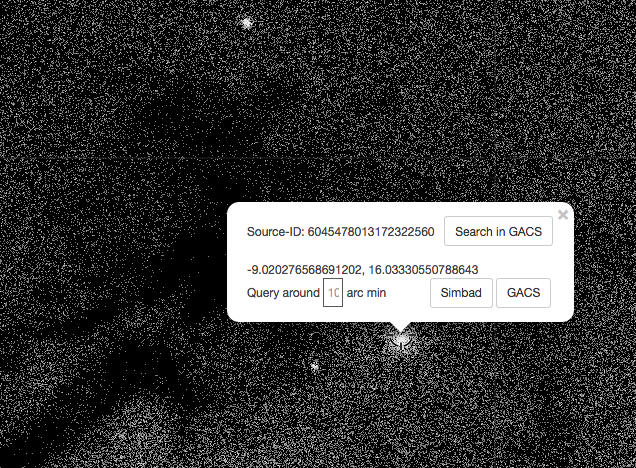}
  \caption{Archive and Simbad searches of object of interest.}\label{fig:simbad_oph}
\end{figure}


Figure~\ref{fig:simbad_oph} illustrates the functionality for archive and Simbad cone searches around objects of interest and also generates an ADQL query-by-identifier for the select object. At any visualisation panel, when the user clicks twice (not double-click) on an object, the  system displays a dialog box with some options. These options, shown in Fig.~\ref{fig:simbad_oph}, identify the selected Source ID from DR1 and give the possibility of generating an ADQL query with this source ID for retrieving further information from the archive. This dialog box also gives the option of retrieving more information from CDS/Simbad or of generating an ADQL cone search query centred on the selected source.

\begin{figure*}[!htbp]
\centering
    \includegraphics[width=0.9\textwidth]{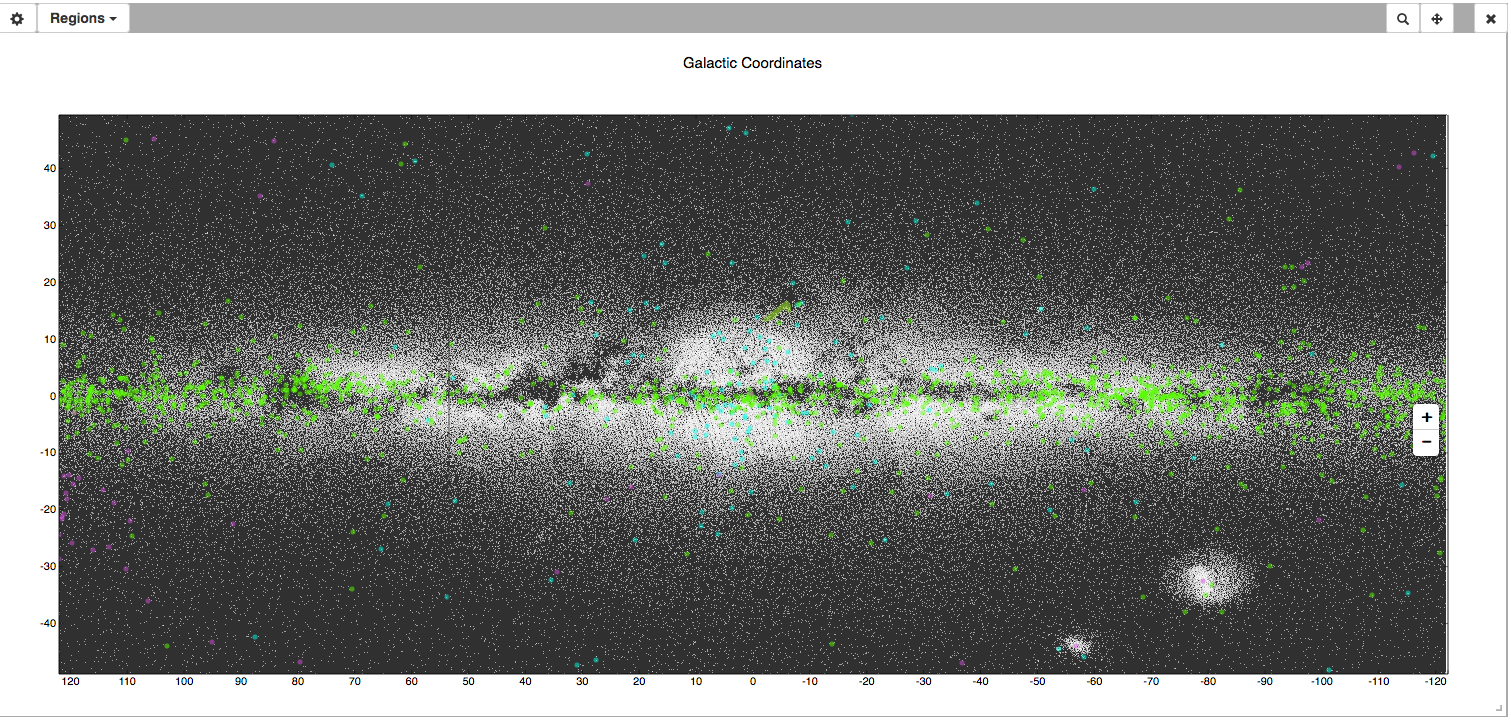}
  \caption{User catalogue of open clusters, globular clusters, and nearby dwarf galaxies overlaid on a scatter plot of the DR1 sources in galactic coordinates.}\label{fig:reg_cat}
\end{figure*}

Figure~\ref{fig:reg_cat} shows catalogues of open clusters \citep[][ Version 3.5, Jan 2016]{2002A&A...389..871D}, globular clusters \citep{1996AJ....112.1487H}, and nearby dwarf galaxies \citep{2012AJ....144....4M} uploaded by the user and overlaid on a scatter plot of the DR1 sources in Galactic coordinates. To upload a catalogue of regions, the user must click on the `Regions' menu, at the top left  of the visualisation window,  then select the `Load Regions' options, and select the region file to be uploaded (we note that the user can save any region  created earlier by using the `Save Regions' option). To overplot points, the region catalogues must be formatted as \emph{region} files with points. An example file can be copied from the region file built from the catalogue of open clusters\footnote{\url{https://gaia.esac.esa.int/gdr1visapp/help/MW_OpenClusters.reg}}.

\section{Other Gaia oriented visualisation tools}
\label{sec:other}

This section provides a small list of applications or services that were identified as having been developed or improved with the exploration of Gaia in mind, and that offer features that complement those of the Archive Visualisation service. It does not intend to be a complete survey of visualisation tools for exploring Gaia data. 

\paragraph{CDS/Aladin} The CDS has an area dedicated to Gaia\footnote{\url{http://cdsweb.u-strasbg.fr/gaia}}. In particular, a page for exploring a DR1 source density HiPS map in Aladin Lite, with optional overlay of individual sources from DR1, TGAS, and SIMBAD is offered\footnote{\url{http://cds.unistra.fr/Gaia/DR1/AL-visualisation.gml}}. The HiPS density map with different HEALPix NSIDE builds can also be downloaded\footnote{\url{http://alasky.u-strasbg.fr/footprints/tables/vizier/I_337_gaia}} .

\paragraph{ESASky}\footnote{\url{http://sky.esa.int}} Gaia catalogues are  available in ESASky \citep{Baines2016}, allowing users to visually compare them with other science catalogues from other ESA missions in an easy way. In this case, users can overplot the Gaia DR1 and TGAS catalogues on top of any image from Gamma-ray to radio wavelengths, click on any single source in the image to identify it in a simplified results table below, and retrieve the selected table as a CSV file or as a VOtable. To cope with potentially slow retrieval times for large fields of view, the resulting table from any search is sorted by median G-magnitude;  only the first 2000 sources are found and it is not possible to select more than 50,000 sources. In the future, ESASky will develop a more sophisticated way to display many sources for large fields of view.

\paragraph{Gaia-Sky}\footnote{\url{https://zah.uni-heidelberg.de/gaia/outreach/gaiasky/}} is a Gaia-focused 3D universe tool intended to run on desktop and laptop systems; its main aim is the production of  outreach material. Gaia-Sky provides a state-of-the-art 3D interactive visualisation of the Gaia catalogue and offers a comprehensive way to visually explain different aspects of the mission.  The latest version contains over 600,000 stars (the stars with relevant parallaxes in TGAS). The  upcoming versions  will be able to display the 1 billion sources of the final Gaia catalogue. The application features different object types such as stars (which can be displayed with their proper motion vectors), planets, moons, asteroids, orbit lines, trajectories, satellites, or constellations. The pace and direction of time can also be tuned interactively via a time warp factor. Graphically, it makes use of advanced rendering techniques and shading algorithms to produce appealing imagery. Internally, the system uses an easily extensible event-driven architecture and is scriptable via Python through a high-level Aplication Programming Interface (API). Different kinds of data sets and objects can also be loaded into the program in a straightforward manner thanks to the simple and human-readable JSON-based format. The system is 3D ready and features four different stereoscopic profiles (cross-eye, parallel view, anaglyphic, and 3DTV); it offers a planetarium mode able to render videos for full-dome systems and a newly added $360{\degr}$ panorama setting which displays the scene in all viewing directions interactively. Gaia Sky is an open source project, it is multi-platform and builds are provided for Linux (RPM, DEB, AUR), Windows (32 and 64 bit versions) and OS X.

\paragraph{TOPCAT}\citep{2005ASPC..347...29T} is a desktop Graphical User Interface (GUI) application for manipulation of source catalogues, widely used to analyse astronomical data\footnote{\url{http://www.star.bris.ac.uk/~mbt/topcat/}}.  One of its features is a large and growing toolkit of highly flexible 1D, 2D, and 3D visualisation options, intended especially for interactive exploration of high-dimensional tabular data. It is suitable for interactive use with hundreds of columns and up to a few million rows on a standard desktop or laptop computer. It can thus work with the whole of the TGAS subset, but not the whole Gaia source catalogue. The visualisation capabilities are also accessible from the corresponding command-line package, STILTS \citep{2006ASPC..351..666T}, which can additionally stream data to generate visualisations from arbitrarily large data sets provided there is  enough computer power.

None of TOPCAT's visualisation capabilities are specific to the Gaia mission, but part of the development work has been carried out within the DPAC, and has accommodated visualisation requirements arising from both preparation and anticipated exploitation of the Gaia catalogue. New features stimulated to date by the requirements of Gaia data analysis include improved control of colour maps; options for assembling, viewing, and exporting HEALPix maps with various aggregation modes; options to view pre-calculated 2D density maps, for instance produced by database queries; improved vector representations, for instance to depict proper motions; plots that trace requested quantiles of noisy data; and Gaussian fitting to histograms. Though developed within the context of Gaia data analysis, all these features are equally applicable to other existing and future data sets.

\paragraph{Vaex}\citep{2016arXiv161204183B} is a visualisation desktop/laptop tool written with the goal of exploring Gaia data\footnote{\url{https://www.astro.rug.nl/~breddels/vaex/}}. It can provide interactive statistical visualisations of over a billion objects in the form of 1D histograms, 2D density plots, and 3D volume renderings. It allows large data volumes to be visualised by computing statistical quantities on a regular grid and displaying visualisations based on those statistics. From the technical point of view, Vaex operates as a HDF5 viewer that exploits the possibilities of memory mapping those files and binning the stored data previous to the rendering and display. However, the full exploration of the over a billion objects requires that the variables of each plot are all loaded in memory. This has the effect of requiring high-end machines, with large amounts of RAM for multi-panel visual exploration of the full DR1. Vaex also operates as a Python library.

\paragraph{Glue}\citep{2015ASPC..495..101B} is a python library for interactive visual data exploration. While not specifically developed for Gaia, a number of uncommon features make Glue deserve a special mention here. It supports the analysis of related data sets spread across different files: a common need of astronomers when analysing data from various sources, including their own observations. A key characteristic is the ability of creating linked views between visualisations of different types of files (images and catalogues).
Glue offers what the authors call \emph{hackable user interfaces}. This means providing GUIs, which are better for interactive  visual exploration, and an API, which is better suited to expressing and automating the creation of visualisations, allowing simple integration in Python notebooks, scripts, and programs. Among other features, Glue also provides advanced capabilities of 3D point cloud selection and support for plug-ins.


\section{Concluding remarks and future developments}
\label{sec:conclusions}

Online, fully interactive, and free visual exploration of the Gaia-sized archives up to the last  of each of the more than $10^9$ individual entries  was something not offered by any service in the world. This scenario has changed with the Gaia Archive Visualisation Service for Data Release 1 presented in this paper. In addition to being used for scientific data exploration and public presentation, the GAVS has also been employed in the validation of DR1 \citep{2017A&A...599A..50A}. 

The software architecture, design, and  implementation have proved highly stable throughout the past months, and is capable of serving a fully interactive visual exploration environment of the Gaia archive to thousands of users.

This work has also been extending the application of visual abstractions of astronomical data sets. It introduces for the first time the simple but powerful concept of a {\it visual query}. This {\it visual query}  directly generates ADQL queries from visual abstractions of the data and tables. It effectively enables any researcher to create complex queries, which can later be executed against ADQL compliant databases such as the one provided by the Gaia Archive. This concept will be evolved in the future to enable even more complex queries, performed through multiple tables, to be built with no knowledge of the ADQL and SQL languages. 

Gaia Data Release 2 will bring a multitude of new parameters. From the astrometric point of view, proper motions, and parallaxes for most of the more than one billion objects will be available. By building on top of concepts and prototypes developed during an exploratory ESA project (code name IVELA: Interactive Visualisation Environment for Large Archives), future versions of the Visualisation Service will provide 3D point cloud interactive visualisation, allowing a fully online 3D navigation and exploration of the release contents.

The future versions of GAVS will also bring useful features such as annotation tools,  various image formats for exportation, and new plot types, such as 2D raster plots (e.g. histograms, density plots) and specialised panels for the analysis of time series, all prepared for very large data sets. Triggering of visualisation pre-computations by users is under analysis. 

Also planned is the extension of GAVS to serve visualisation data (indexes, levels of detail, linked views, and more metadata) to other applications beyond the the current web client/portal. The baseline is currently a REST API, with wrappers planned for python and other languages.
Advanced data analysis will then become possible with tools (e.g. Glue) that otherwise would not be able to handle the volume of the Gaia Archive, but the data feeds do not have to be limited to data analysis frameworks. The success of DR1 has demonstrated the high level of  interest among the general public in the Gaia mission. It would thus also  be natural to  feed education and outreach Universe exploration tools such as Gaia-Sky or the World Wide Telescope\footnote{\url{http://www.worldwidetelescope.org/}}.

Finally, further ahead, a future which includes a deeper articulation with virtual organisations such as the Virtual Observatory seems unavoidable. In the light of this paradigm, it is only  natural that code should be brought close to the data, and not the other way around. Accordingly, there have been studies, designs, and developments of platforms such as the  Gaia Added Value Platform (GAVIP) \citep{GAVIP2016}, which would allow codes, for instance  Python, to run near the Gaia data. The Visualisation functionality of the GAVIP platform, known as the Gaia Added Value Interface for Data Visualisation (GAVIDAV), has been developed in close contact with the Gaia Archive Visualisation Services. This proximity will enable any application running on such a platform, and thus near the Gaia Archive, to profit from many of the large-data visualisation capabilities of the tools described in this paper, thus bringing the power of visually exploring billions of database entries to the hands of {\it any} astronomer or human being, regardless of the levels of resources available in  their country or institution.

\begin{acknowledgements}
The authors greatly appreciated the constructive comments by the referee, Alyssa Goodman. 
This work has made use of results from the European Space Agency (ESA) space mission Gaia, whose data were processed by the Gaia Data Processing and Analysis Consortium (DPAC). Funding for the DPAC has been provided by national institutions, in particular the institutions participating in the Gaia Multilateral Agreement. The Gaia mission website is \url{http: //www.cosmos.esa.int/gaia}. The authors are current or past members of the ESA Gaia mission team and of the Gaia DPAC. This work has received financial support from the European Commission’s Seventh Framework Programme through the grant FP7-606740 (FP7-SPACE-2013-1) for the Gaia European Network for Improved data User Services (GENIUS); from the Portuguese Funda\c c\~ao para a Ci\^encia e a Tecnologia (FCT) through grants PTDC/CTE-SPA/118692/2010, PDCTE/CTE-AST/81711/2003, and SFRH/BPD/74697/2010; from the Portuguese Strategic Programmes PEstOE/AMB/UI4006/2011 for SIM, UID/FIS/00099/2013 for CENTRA, and UID/EEA/00066/2013 for UNINOVA; from the ESA contracts ESA/RFQ/3-14211/14/NL/JD and ITT-AO/1-7094/12/NL/CO Ref:B00015862. This research has made use of the Set of Identifications, Measurements, and Bibliography for Astronomical Data \citep{Wenger2000} and of the ``Aladin sky atlas'' \citep{2000A&AS..143...33B,2014ASPC..485..277B}, which are developed and operated at Centre de Donn\'ees astronomiques de Strasbourg (CDS), France. 
XL acknowledges support by the MINECO (Spanish Ministry of Economy) - FEDER through grant ESP2014-55996-C2-1-R,  MDM-2014-0369 of ICCUB (Unidad de Excelencia `Mar\'ia de Maeztu').
This publication made use of data products from the Two Micron All Sky Survey, which is a joint project of the University of Massachusetts and the Infrared Processing and Analysis Center/California Institute of Technology, funded by the National Aeronautics and Space Administration and the National Science Foundation. Thomas Boch and Eric Mandel are warmly thanked for their always friendly assistance in integrating the Aladin Lite and JS9 plug-ins, respectively.
      
\end{acknowledgements}


\begin{thebibliography}{46}
\expandafter\ifx\csname natexlab\endcsname\relax\def\natexlab#1{#1}\fi

\bibitem[{{Arenou} {et~al.}(2017){Arenou}, {Luri}, {Babusiaux}, {Fabricius},
  {Helmi}, {Robin}, {Vallenari}, {Blanco-Cuaresma}, {Cantat-Gaudin},
  {Findeisen}, {Reyl{\'e}}, {Ruiz-Dern}, {Sordo}, {Turon}, {Walton}, {Shih},
  {Antiche}, {Barache}, {Barros}, {Breddels}, {Carrasco}, {Costigan},
  {Diakit{\'e}}, {Eyer}, {Figueras}, {Galluccio}, {Heu}, {Jordi},
  {Krone-Martins}, {Lallement}, {Lambert}, {Leclerc}, {Marrese}, {Moitinho},
  {Mor}, {Romero-G{\'o}mez}, {Sartoretti}, {Soria}, {Soubiran}, {Souchay},
  {Veljanoski}, {Ziaeepour}, {Giuffrida}, {Pancino}, \&
  {Bragaglia}}]{2017A&A...599A..50A}
{Arenou}, F., {Luri}, X., {Babusiaux}, C., {et~al.} 2017, \aap, 599, A50

\bibitem[{{Bailer-Jones} {et~al.}(2013){Bailer-Jones}, {Andrae}, {Arcay},
  {Astraatmadja}, {Bellas-Velidis}, {Berihuete}, {Bijaoui}, {Carri{\'o}n},
  {Dafonte}, {Damerdji}, {Dapergolas}, {de Laverny}, {Delchambre}, {Drazinos},
  {Drimmel}, {Fr{\'e}mat}, {Fustes}, {Garc{\'{\i}}a-Torres}, {Gu{\'e}d{\'e}},
  {Heiter}, {Janotto}, {Karampelas}, {Kim}, {Knude}, {Kolka}, {Kontizas},
  {Kontizas}, {Korn}, {Lanzafame}, {Lebreton}, {Lindstr{\o}m}, {Liu},
  {Livanou}, {Lobel}, {Manteiga}, {Martayan}, {Ordenovic}, {Pichon},
  {Recio-Blanco}, {Rocca-Volmerange}, {Sarro}, {Smith}, {Sordo}, {Soubiran},
  {Surdej}, {Th{\'e}venin}, {Tsalmantza}, {Vallenari}, \&
  {Zorec}}]{2013A&A...559A..74B}
{Bailer-Jones}, C.~A.~L., {Andrae}, R., {Arcay}, B., {et~al.} 2013, \aap, 559,
  A74

\bibitem[{{Baines} {et~al.}(2017{\natexlab{a}}){Baines}, {Giordano}, {Racero},
  {Salgado}, {L{\'o}pez Mart{\'{\i}}}, {Mer{\'{\i}}n}, {Sarmiento},
  {Guti{\'e}rrez}, {Ortiz de Landaluce}, {Le{\'o}n}, {de Teodoro},
  {Gonz{\'a}lez}, {Nieto}, {Segovia}, {Pollock}, {Rosa}, {Arviset}, {Lennon},
  {O'Mullane}, \& {de Marchi}}]{2017PASP..129b8001B}
{Baines}, D., {Giordano}, F., {Racero}, E., {et~al.} 2017{\natexlab{a}}, \pasp,
  129, 028001

\bibitem[{{Baines} {et~al.}(2017{\natexlab{b}}){Baines}, {Giordano}, {Racero},
  {Salgado}, {L{\'o}pez Mart{\'{\i}}}, {Mer{\'{\i}}n}, {Sarmiento},
  {Guti{\'e}rrez}, {Ortiz de Landaluce}, {Le{\'o}n}, {de Teodoro},
  {Gonz{\'a}lez}, {Nieto}, {Segovia}, {Pollock}, {Rosa}, {Arviset}, {Lennon},
  {O'Mullane}, \& {de Marchi}}]{Baines2016}
{Baines}, D., {Giordano}, F., {Racero}, E., {et~al.} 2017{\natexlab{b}}, \pasp,
  129, 028001

\bibitem[{{Beaumont} {et~al.}(2015){Beaumont}, {Goodman}, \&
  {Greenfield}}]{2015ASPC..495..101B}
{Beaumont}, C., {Goodman}, A., \& {Greenfield}, P. 2015, in Astronomical
  Society of the Pacific Conference Series, Vol. 495, Astronomical Data
  Analysis Software an Systems XXIV (ADASS XXIV), ed. A.~R. {Taylor} \&
  E.~{Rosolowsky}, 101

\bibitem[{{Boch} \& {Fernique}(2014)}]{2014ASPC..485..277B}
{Boch}, T. \& {Fernique}, P. 2014, in Astronomical Society of the Pacific
  Conference Series, Vol. 485, Astronomical Data Analysis Software and Systems
  XXIII, ed. N.~{Manset} \& P.~{Forshay}, 277

\bibitem[{{Bonnarel} {et~al.}(2000){Bonnarel}, {Fernique}, {Bienaym{\'e}},
  {Egret}, {Genova}, {Louys}, {Ochsenbein}, {Wenger}, \&
  {Bartlett}}]{2000A&AS..143...33B}
{Bonnarel}, F., {Fernique}, P., {Bienaym{\'e}}, O., {et~al.} 2000, \aaps, 143,
  33

\bibitem[{{Breddels}(2016)}]{2016arXiv161204183B}
{Breddels}, M.~A. 2016, ArXiv e-prints [\eprint[arXiv]{1612.04183}]

\bibitem[{{Brown} {et~al.}(2012){Brown}, {Arenou}, {Hambly}, {van Leeuwen},
  {Luri}, {Malapert}, {O'Mullane}, {Tapiador}, \& {Walton}}]{BrownTN026}
{Brown}, A., {Arenou}, F., {Hambly}, N., {et~al.} 2012, Gaia DPAC Technical
  Note GAIA-C9-TN-LEI-AB-026

\bibitem[{{Dawson} {et~al.}(2016){Dawson}, {Kneib}, {Percival}, {Alam},
  {Albareti}, {Anderson}, {Armengaud}, {Aubourg}, {Bailey}, {Bautista},
  {Berlind}, {Bershady}, {Beutler}, {Bizyaev}, {Blanton}, {Blomqvist},
  {Bolton}, {Bovy}, {Brandt}, {Brinkmann}, {Brownstein}, {Burtin}, {Busca},
  {Cai}, {Chuang}, {Clerc}, {Comparat}, {Cope}, {Croft}, {Cruz-Gonzalez}, {da
  Costa}, {Cousinou}, {Darling}, {de la Macorra}, {de la Torre}, {Delubac}, {du
  Mas des Bourboux}, {Dwelly}, {Ealet}, {Eisenstein}, {Eracleous}, {Escoffier},
  {Fan}, {Finoguenov}, {Font-Ribera}, {Frinchaboy}, {Gaulme}, {Georgakakis},
  {Green}, {Guo}, {Guy}, {Ho}, {Holder}, {Huehnerhoff}, {Hutchinson}, {Jing},
  {Jullo}, {Kamble}, {Kinemuchi}, {Kirkby}, {Kitaura}, {Klaene}, {Laher},
  {Lang}, {Laurent}, {Le Goff}, {Li}, {Liang}, {Lima}, {Lin}, {Lin}, {Lin},
  {Long}, {Lundgren}, {MacDonald}, {Geimba Maia}, {Malanushenko},
  {Malanushenko}, {Mariappan}, {McBride}, {McGreer}, {M{\'e}nard}, {Merloni},
  {Meza}, {Montero-Dorta}, {Muna}, {Myers}, {Nandra}, {Naugle}, {Newman},
  {Noterdaeme}, {Nugent}, {Ogando}, {Olmstead}, {Oravetz}, {Oravetz},
  {Padmanabhan}, {Palanque-Delabrouille}, {Pan}, {Parejko}, {P{\^a}ris},
  {Peacock}, {Petitjean}, {Pieri}, {Pisani}, {Prada}, {Prakash}, {Raichoor},
  {Reid}, {Rich}, {Ridl}, {Rodriguez-Torres}, {Carnero Rosell}, {Ross},
  {Rossi}, {Ruan}, {Salvato}, {Sayres}, {Schneider}, {Schlegel}, {Seljak},
  {Seo}, {Sesar}, {Shandera}, {Shu}, {Slosar}, {Sobreira}, {Streblyanska},
  {Suzuki}, {Taylor}, {Tao}, {Tinker}, {Tojeiro}, {Vargas-Maga{\~n}a}, {Wang},
  {Weaver}, {Weinberg}, {White}, {Wood-Vasey}, {Yeche}, {Zhai}, {Zhao}, {Zhao},
  {Zheng}, {Ben Zhu}, \& {Zou}}]{2016AJ....151...44D}
{Dawson}, K.~S., {Kneib}, J.-P., {Percival}, W.~J., {et~al.} 2016, \aj, 151, 44

\bibitem[{{Dias} {et~al.}(2002){Dias}, {Alessi}, {Moitinho}, \&
  {L{\'e}pine}}]{2002A&A...389..871D}
{Dias}, W.~S., {Alessi}, B.~S., {Moitinho}, A., \& {L{\'e}pine}, J.~R.~D. 2002,
  \aap, 389, 871

\bibitem[{Ellis \& Dix(2007)}]{4376143}
Ellis, G. \& Dix, A. 2007, IEEE Transactions on Visualization and Computer
  Graphics, 13, 1216

\bibitem[{{Eyer} {et~al.}(2014){Eyer}, {Evans}, {Mowlavi}, {Lanzafame},
  {Cuypers}, {De Ridder}, {Sarro}, {Clementini}, {Guy}, {Holl}, {Ordonez},
  {Nienartowicz}, \& {Lecoeur-Taibi}}]{2014EAS....67...75E}
{Eyer}, L., {Evans}, D.~W., {Mowlavi}, N., {et~al.} 2014, in EAS Publications
  Series, Vol.~67, EAS Publications Series, 75--78

\bibitem[{Fielding(2000)}]{Fielding:2000}
Fielding, R.~T. 2000, PhD thesis, university of California, Irvine, AAI9980887

\bibitem[{{Gaia Collaboration} {et~al.}(2016{\natexlab{a}}){Gaia
  Collaboration}, {Brown}, {Vallenari}, {Prusti}, {de Bruijne}, {Mignard},
  {Drimmel}, {Babusiaux}, {Bailer-Jones}, {Bastian}, \&
  et~al.}]{2016A&A...595A...2G}
{Gaia Collaboration}, {Brown}, A.~G.~A., {Vallenari}, A., {et~al.}
  2016{\natexlab{a}}, \aap, 595, A2

\bibitem[{{Gaia Collaboration} {et~al.}(2016{\natexlab{b}}){Gaia
  Collaboration}, {Prusti}, {de Bruijne}, {Brown}, {Vallenari}, {Babusiaux},
  {Bailer-Jones}, {Bastian}, {Biermann}, {Evans}, \&
  et~al.}]{2016A&A...595A...1G}
{Gaia Collaboration}, {Prusti}, T., {de Bruijne}, J.~H.~J., {et~al.}
  2016{\natexlab{b}}, \aap, 595, A1

\bibitem[{{Goodman}(2012)}]{2012AN....333..505G}
{Goodman}, A.~A. 2012, Astronomische Nachrichten, 333, 505

\bibitem[{Gorski {et~al.}(2005)Gorski, Hivon, Banday, Wandelt, Hansen,
  Reinecke, \& Bartelmann}]{gorski2005healpix}
Gorski, K.~M., Hivon, E., Banday, A., {et~al.} 2005, The Astrophysical Journal,
  622, 759

\bibitem[{{Harris}(1996)}]{1996AJ....112.1487H}
{Harris}, W.~E. 1996, \aj, 112, 1487

\bibitem[{{Hassan} {et~al.}(2013){Hassan}, {Fluke}, {Barnes}, \&
  {Kilborn}}]{2013MNRAS.429.2442H}
{Hassan}, A.~H., {Fluke}, C.~J., {Barnes}, D.~G., \& {Kilborn}, V.~A. 2013,
  \mnras, 429, 2442

\bibitem[{Hey {et~al.}(2009)Hey, Tansley, \& Tolle}]{hey:fourthparadigm:2009}
Hey, T., Tansley, S., \& Tolle, K. 2009, The Fourth Paradigm: Data-Intensive
  Scientific Discovery (Microsoft Research)

\bibitem[{{Ivezic} {et~al.}(2008){Ivezic}, {Tyson}, {Abel}, {Acosta},
  {Allsman}, {AlSayyad}, {Anderson}, {Andrew}, {Angel}, {Angeli}, {Ansari},
  {Antilogus}, {Arndt}, {Astier}, {Aubourg}, {Axelrod}, {Bard}, {Barr},
  {Barrau}, {Bartlett}, {Bauman}, {Beaumont}, {Becker}, {Becla}, {Beldica},
  {Bellavia}, {Blanc}, {Blandford}, {Bloom}, {Bogart}, {Borne}, {Bosch},
  {Boutigny}, {Brandt}, {Brown}, {Bullock}, {Burchat}, {Burke}, {Cagnoli},
  {Calabrese}, {Chandrasekharan}, {Chesley}, {Cheu}, {Chiang}, {Claver},
  {Connolly}, {Cook}, {Cooray}, {Covey}, {Cribbs}, {Cui}, {Cutri}, {Daubard},
  {Daues}, {Delgado}, {Digel}, {Doherty}, {Dubois}, {Dubois-Felsmann},
  {Durech}, {Eracleous}, {Ferguson}, {Frank}, {Freemon}, {Gangler}, {Gawiser},
  {Geary}, {Gee}, {Geha}, {Gibson}, {Gilmore}, {Glanzman}, {Goodenow},
  {Gressler}, {Gris}, {Guyonnet}, {Hascall}, {Haupt}, {Hernandez}, {Hogan},
  {Huang}, {Huffer}, {Innes}, {Jacoby}, {Jain}, {Jee}, {Jernigan},
  {Jevremovic}, {Johns}, {Jones}, {Juramy-Gilles}, {Juric}, {Kahn}, {Kalirai},
  {Kallivayalil}, {Kalmbach}, {Kantor}, {Kasliwal}, {Kessler}, {Kirkby},
  {Knox}, {Kotov}, {Krabbendam}, {Krughoff}, {Kubanek}, {Kuczewski},
  {Kulkarni}, {Lambert}, {Le Guillou}, {Levine}, {Liang}, {Lim}, {Lintott},
  {Lupton}, {Mahabal}, {Marshall}, {Marshall}, {May}, {McKercher}, {Migliore},
  {Miller}, {Mills}, {Monet}, {Moniez}, {Neill}, {Nief}, {Nomerotski},
  {Nordby}, {O'Connor}, {Oliver}, {Olivier}, {Olsen}, {Ortiz}, {Owen}, {Pain},
  {Peterson}, {Petry}, {Pierfederici}, {Pietrowicz}, {Pike}, {Pinto}, {Plante},
  {Plate}, {Price}, {Prouza}, {Radeka}, {Rajagopal}, {Rasmussen}, {Regnault},
  {Ridgway}, {Ritz}, {Rosing}, {Roucelle}, {Rumore}, {Russo}, {Saha},
  {Sassolas}, {Schalk}, {Schindler}, {Schneider}, {Schumacher}, {Sebag},
  {Sembroski}, {Seppala}, {Shipsey}, {Silvestri}, {Smith}, {Smith}, {Strauss},
  {Stubbs}, {Sweeney}, {Szalay}, {Takacs}, {Thaler}, {Van Berg}, {Vanden Berk},
  {Vetter}, {Virieux}, {Xin}, {Walkowicz}, {Walter}, {Wang}, {Warner},
  {Willman}, {Wittman}, {Wolff}, {Wood-Vasey}, {Yoachim}, {Zhan}, \& {for the
  LSST Collaboration}}]{2008arXiv0805.2366I}
{Ivezic}, Z., {Tyson}, J.~A., {Abel}, B., {et~al.} 2008, ArXiv e-prints
  [\eprint[arXiv]{0805.2366}]

\bibitem[{Jern {et~al.}(2007)Jern, Johansson, Johansson, \&
  Franzen}]{Jern:2007}
Jern, M., Johansson, S., Johansson, J., \& Franzen, J. 2007, in Proceedings of
  the Fifth International Conference on Coordinated and Multiple Views in
  Exploratory Visualization, CMV '07 (Washington, DC, USA: IEEE Computer
  Society), 85--97

\bibitem[{{Katz} {et~al.}(2011){Katz}, {Cropper}, {Meynadier}, {Jean-Antoine},
  {Allende Prieto}, {Baker}, {Benson}, {Berthier}, {Bigot}, {Blomme},
  {Boudreault}, {Chemin}, {Crifo}, {Damerdji}, {David}, {David}, {Delle Luche},
  {Dolding}, {Fr{\'e}mat}, {Gerbier}, {Gerssen}, {G{\'o}mez}, {Gosset},
  {Guerrier}, {Guy}, {Hall}, {Hestroffer}, {Huckle}, {Jasniewicz}, {Ludwig},
  {Martayan}, {Morel}, {Nguyen}, {Ocvirk}, {Parr}, {Royer}, {Sartoretti},
  {Seabroke}, {Simon}, {Smith}, {Soubiran}, {Steinmetz}, {Th{\'e}venin},
  {Turon}, {Udry}, {Veltz}, \& {Viala}}]{2011EAS....45..189K}
{Katz}, D., {Cropper}, M., {Meynadier}, F., {et~al.} 2011, in EAS Publications
  Series, Vol.~45, EAS Publications Series, ed. C.~{Turon}, F.~{Meynadier}, \&
  F.~{Arenou}, 189--194

\bibitem[{Keim(2002)}]{Keim:2002:IVV:614285.614508}
Keim, D.~A. 2002, IEEE Transactions on Visualization and Computer Graphics, 8,
  1

\bibitem[{{Krone-Martins} {et~al.}(2013){Krone-Martins}, {Ducourant},
  {Teixeira}, {Galluccio}, {Gavras}, {dos Anjos}, {de Souza}, {Machado}, \& {Le
  Campion}}]{2013A&A...556A.102K}
{Krone-Martins}, A., {Ducourant}, C., {Teixeira}, R., {et~al.} 2013, \aap, 556,
  A102

\bibitem[{{Laureijs} {et~al.}(2011){Laureijs}, {Amiaux}, {Arduini},
  {Augu{\`e}res}, {Brinchmann}, {Cole}, {Cropper}, {Dabin}, {Duvet}, {Ealet},
  \& et~al.}]{2011arXiv1110.3193L}
{Laureijs}, R., {Amiaux}, J., {Arduini}, S., {et~al.} 2011, ArXiv e-prints
  [\eprint[arXiv]{1110.3193}]

\bibitem[{{Lindegren} {et~al.}(2016){Lindegren}, {Lammers}, {Bastian},
  {Hern{\'a}ndez}, {Klioner}, {Hobbs}, {Bombrun}, {Michalik}, {Ramos-Lerate},
  {Butkevich}, {Comoretto}, {Joliet}, {Holl}, {Hutton}, {Parsons},
  {Steidelm{\"u}ller}, {Abbas}, {Altmann}, {Andrei}, {Anton}, {Bach},
  {Barache}, {Becciani}, {Berthier}, {Bianchi}, {Biermann}, {Bouquillon},
  {Bourda}, {Br{\"u}semeister}, {Bucciarelli}, {Busonero}, {Carlucci},
  {Casta{\~n}eda}, {Charlot}, {Clotet}, {Crosta}, {Davidson}, {de Felice},
  {Drimmel}, {Fabricius}, {Fienga}, {Figueras}, {Fraile}, {Gai}, {Garralda},
  {Geyer}, {Gonz{\'a}lez-Vidal}, {Guerra}, {Hambly}, {Hauser}, {Jordan},
  {Lattanzi}, {Lenhardt}, {Liao}, {L{\"o}ffler}, {McMillan}, {Mignard}, {Mora},
  {Morbidelli}, {Portell}, {Riva}, {Sarasso}, {Serraller}, {Siddiqui}, {Smart},
  {Spagna}, {Stampa}, {Steele}, {Taris}, {Torra}, {van Reeven}, {Vecchiato},
  {Zschocke}, {de Bruijne}, {Gracia}, {Raison}, {Lister}, {Marchant},
  {Messineo}, {Soffel}, {Osorio}, {de Torres}, \&
  {O'Mullane}}]{2016A&A...595A...4L}
{Lindegren}, L., {Lammers}, U., {Bastian}, U., {et~al.} 2016, \aap, 595, A4

\bibitem[{{McConnachie}(2012)}]{2012AJ....144....4M}
{McConnachie}, A.~W. 2012, \aj, 144, 4

\bibitem[{{Ortiz} {et~al.}(2008){Ortiz}, {Lusted}, {Dowler}, {Szalay},
  {Shirasaki}, {NietoSantisteban}, {Ohishi}, {O'Mullane}, \&
  {Osuna}}]{2008IVOAADQL}
{Ortiz}, I., {Lusted}, J., {Dowler}, P., {et~al.} 2008, IVOA Recommendation on
  ADQL 2.0

\bibitem[{Peng {et~al.}(2004)Peng, Ward, \&
  Rundensteiner}]{Peng04clutterreduction}
Peng, W., Ward, M.~O., \& Rundensteiner, E.~A. 2004, IEEE InfoVis, 89

\bibitem[{{Pourbaix}(2011)}]{2011AIPC.1346..122P}
{Pourbaix}, D. 2011, in American Institute of Physics Conference Series, Vol.
  1346, American Institute of Physics Conference Series, ed. J.~A. {Docobo},
  V.~S. {Tamazian}, \& Y.~Y. {Balega}, 122--133

\bibitem[{Rosenholtz {et~al.}(2005)Rosenholtz, Li, Mansfield, \&
  Jin}]{Rosenholtz05featurecongestion}
Rosenholtz, R., Li, Y., Mansfield, J., \& Jin, Z. 2005, in Proc. ACM SIGCHI
  Conference on Human Factors in Computing Systems, 761--770

\bibitem[{Synder(1993)}]{Synder1993}
Synder, J.~P. 1993, Flattening the Earth: Two Thousand Years of Map Projections
  (University of Chicago Press)

\bibitem[{{Szalay} \& {Gray}(2001)}]{2001Sci...293.2037S}
{Szalay}, A. \& {Gray}, J. 2001, Science, 293, 2037

\bibitem[{{Szalay} {et~al.}(2008){Szalay}, {Springel}, \&
  {Lemson}}]{2008arXiv0811.2055S}
{Szalay}, T., {Springel}, V., \& {Lemson}, G. 2008, ArXiv e-prints
  [\eprint[arXiv]{0811.2055}]

\bibitem[{Tanaka(2014)}]{Tanaka:2014}
Tanaka, Y. 2014, in Proc. of the 18th Int. Conf. on Inf. Vis., IV ’14,
  170--175

\bibitem[{{Tanga} {et~al.}(2016){Tanga}, {Mignard}, {Dell`Oro}, {Muinonen},
  {Pauwels}, {Thuillot}, {Berthier}, {Cellino}, {Hestroffer}, {Petit}, {Carry},
  {David}, {Delbo`}, {Fedorets}, {Galluccio}, {Granvik}, {Ordenovic}, \&
  {Pentik{\"a}inen}}]{2016P&SS..123...87T}
{Tanga}, P., {Mignard}, F., {Dell`Oro}, A., {et~al.} 2016, \planss, 123, 87

\bibitem[{{Taylor}(2005)}]{2005ASPC..347...29T}
{Taylor}, M.~B. 2005, in Astronomical Society of the Pacific Conference Series,
  Vol. 347, Astronomical Data Analysis Software and Systems XIV, ed.
  P.~{Shopbell}, M.~{Britton}, \& R.~{Ebert}, 29

\bibitem[{{Taylor}(2006)}]{2006ASPC..351..666T}
{Taylor}, M.~B. 2006, in Astronomical Society of the Pacific Conference Series,
  Vol. 351, Astronomical Data Analysis Software and Systems XV, ed.
  C.~{Gabriel}, C.~{Arviset}, D.~{Ponz}, \& S.~{Enrique}, 666

\bibitem[{{Tukey}(1977)}]{tukey1977}
{Tukey}, J.~W. 1977, {Exploratory Data Analysis}, ed. {Tukey, J.~W.},
  Behavioral Science: Quantitative Methods (Reading, Mass.: Addison-Wesley)

\bibitem[{Unwin {et~al.}(2006)Unwin, Theus, \& Hofmann}]{unwin:graphics:2006}
Unwin, A., Theus, M., \& Hofmann, H. 2006, {Graphics of Large Datasets:
  Visualizing a Million}, 1st edn., Statistics and Computing (Springer)

\bibitem[{Vagg {et~al.}(2016)Vagg, O'Callaghan, \'O~h\'Og\'ain, McBreen,
  Hanlon, Lynn, \& O'Mullane}]{GAVIP2016}
Vagg, D., O'Callaghan, D., \'O~h\'Og\'ain, F., {et~al.} 2016, Proc. SPIE, 9913
  [\eprint{https://arxiv.org/pdf/1605.09287v1.pdf}]

\bibitem[{{van Leeuwen} {et~al.}(2017){van Leeuwen}, {Evans}, {De Angeli},
  {Jordi}, {Busso}, {Cacciari}, {Riello}, {Pancino}, {Altavilla}, {Brown},
  {Burgess}, {Carrasco}, {Cocozza}, {Cowell}, {Davidson}, {De Luise},
  {Fabricius}, {Galleti}, {Gilmore}, {Giuffrida}, {Hambly}, {Harrison},
  {Hodgkin}, {Holland}, {MacDonald}, {Marinoni}, {Montegriffo}, {Osborne},
  {Ragaini}, {Richards}, {Rowell}, {Voss}, {Walton}, {Weiler}, {Castellani},
  {Delgado}, {H{\o}g}, {van Leeuwen}, {Millar}, {Pagani}, {Piersimoni},
  {Pulone}, {Rixon}, {Suess}, {Wyrzykowski}, {Yoldas}, {Alecu}, {Allan},
  {Balaguer-N{\'u}{\~n}ez}, {Barstow}, {Bellazzini}, {Belokurov},
  {Blagorodnova}, {Bonfigli}, {Bragaglia}, {Brown}, {Bunclark}, {Buonanno},
  {Burgon}, {Campbell}, {Collins}, {Cross}, {Ducourant}, {van Elteren},
  {Evans}, {Federici}, {Fern{\'a}ndez-Hern{\'a}ndez}, {Figueras}, {Fraser},
  {Fyfe}, {Gebran}, {Heyrovsky}, {Holl}, {Holland}, {Iannicola}, {Irwin},
  {Koposov}, {Krone-Martins}, {Mann}, {Marrese}, {Masana}, {Munari}, {Ortiz},
  {Ouzounis}, {Peltzer}, {Portell}, {Read}, {Terrett}, {Torra}, {Trager},
  {Troisi}, {Valentini}, {Vallenari}, \& {Wevers}}]{2017A&A...599A..32V}
{van Leeuwen}, F., {Evans}, D.~W., {De Angeli}, F., {et~al.} 2017, \aap, 599,
  A32

\bibitem[{{Wenger} {et~al.}(2000){Wenger}, {Ochsenbein}, {Egret}, {Dubois},
  {Bonnarel}, {Borde}, {Genova}, {Jasniewicz}, {Lalo{\"e}}, {Lesteven}, \&
  {Monier}}]{Wenger2000}
{Wenger}, M., {Ochsenbein}, F., {Egret}, D., {et~al.} 2000, \aaps, 143, 9

\bibitem[{{York} {et~al.}(2000){York}, {Adelman}, {Anderson}, {Anderson},
  {Annis}, {Bahcall}, {Bakken}, {Barkhouser}, {Bastian}, {Berman}, {Boroski},
  {Bracker}, {Briegel}, {Briggs}, {Brinkmann}, {Brunner}, {Burles}, {Carey},
  {Carr}, {Castander}, {Chen}, {Colestock}, {Connolly}, {Crocker}, {Csabai},
  {Czarapata}, {Davis}, {Doi}, {Dombeck}, {Eisenstein}, {Ellman}, {Elms},
  {Evans}, {Fan}, {Federwitz}, {Fiscelli}, {Friedman}, {Frieman}, {Fukugita},
  {Gillespie}, {Gunn}, {Gurbani}, {de Haas}, {Haldeman}, {Harris}, {Hayes},
  {Heckman}, {Hennessy}, {Hindsley}, {Holm}, {Holmgren}, {Huang}, {Hull},
  {Husby}, {Ichikawa}, {Ichikawa}, {Ivezi{\'c}}, {Kent}, {Kim}, {Kinney},
  {Klaene}, {Kleinman}, {Kleinman}, {Knapp}, {Korienek}, {Kron}, {Kunszt},
  {Lamb}, {Lee}, {Leger}, {Limmongkol}, {Lindenmeyer}, {Long}, {Loomis},
  {Loveday}, {Lucinio}, {Lupton}, {MacKinnon}, {Mannery}, {Mantsch}, {Margon},
  {McGehee}, {McKay}, {Meiksin}, {Merelli}, {Monet}, {Munn}, {Narayanan},
  {Nash}, {Neilsen}, {Neswold}, {Newberg}, {Nichol}, {Nicinski}, {Nonino},
  {Okada}, {Okamura}, {Ostriker}, {Owen}, {Pauls}, {Peoples}, {Peterson},
  {Petravick}, {Pier}, {Pope}, {Pordes}, {Prosapio}, {Rechenmacher}, {Quinn},
  {Richards}, {Richmond}, {Rivetta}, {Rockosi}, {Ruthmansdorfer}, {Sandford},
  {Schlegel}, {Schneider}, {Sekiguchi}, {Sergey}, {Shimasaku}, {Siegmund},
  {Smee}, {Smith}, {Snedden}, {Stone}, {Stoughton}, {Strauss}, {Stubbs},
  {SubbaRao}, {Szalay}, {Szapudi}, {Szokoly}, {Thakar}, {Tremonti}, {Tucker},
  {Uomoto}, {Vanden Berk}, {Vogeley}, {Waddell}, {Wang}, {Watanabe},
  {Weinberg}, {Yanny}, {Yasuda}, \& {SDSS Collaboration}}]{2000AJ....120.1579Y}
{York}, D.~G., {Adelman}, J., {Anderson}, Jr., J.~E., {et~al.} 2000, \aj, 120,
  1579

\end{thebibliography}



\section{Appendix A -- Gaia Data Access Scenarios related to visualisation}

This appendix lists a subset of the Gaia Data Access Scenarios \citep{BrownTN026}\footnote{The document can be accessed at \url{http://www.rssd.esa.int/doc\_fetch.php?id=3125400}
} that are related to visualisation and that were used to drive the design of the system presented in this paper. 

\begin{itemize}
\item{\bf GDAS-BR-10}: I would like to be able to select objects based on any set of the variables provided in the Gaia catalogue position, parallax, astrophysical parameters, proper motion uncertainties, etc. These selections should not be limited to simple 'axis-parallel' cuts or cone cuts, but permit a broader array of functions/functional dependencies. An example is selection of fractional parallax error and some relation between G magnitude and extinction. Another is selection on space velocities, which requires a combination of position, parallax, and proper motion.

\item{\bf GDAS-BR-07}: I want data of all objects contained in a rectangular/circular region centred on a given sky position.

\item{\bf GDAS-GA-02}: I would like to query the Gaia catalogue in some peculiar Galactic directions and retrieve the (U, V, W) velocities of the stars relative to the Sun when possible, together with their distance, metallicity, and error bars.

\item{\bf GDAS-ST-25}: I want to display the information for a given source (for variability analysis): source attributes (e.g. mean magnitude, mean colour, period, amplitude, etc.), light curve, folded light curve, frequency gramme.

\item{\bf GDAS-ST-06}: Proximity Queries: I would like to select targets in rectangle, polygon, ellipse; select objects in this circular part of the sky that are closer to target X than to target Y; coordinate systems: equatorial, ecliptic, galactic.

\item{\bf GDAS-EG-08}: I want all quasars observed so far with Gaia to be plotted on the celestial sphere.

\item{\bf GDAS-GA-15}: I would like to retrieve the astrometric and photometric data concerning any star hosting planets. I need positions, proper motions, parallaxes, and radial velocities (if available) and the associated covariance matrices.

\item{\bf GDAS-BR-16}: Interactive plots (2D/3D zooming and rotation). I would like to have the possibility to select any attribute for any axis and customise the plot in that way; e.g. selecting mean values for colour and magnitudes, the user can produce the HR diagram. It should be possible to make a selection of points and store the set of sources for other plots or analysis. Aggregated plots: when the number of objects is huge, the data has to be aggregated for better visualisation. Kinds of aggregation plots include: IQR plot, Histogram, Mosaic plots, Bubble plots, Density plots. These aggregated plots are meant to analyse large sets of data without plotting detailed information for all sources.

\item{\bf GDAS-BR-02}: Show me the area of sky which Gaia will be observing on any particular date.

\item{\bf GDAS-GA-04}: I want whatever measurements of MW G2 stars only in a given sky region.

\item{\bf GDAS-PR-03}:  I want to make a movie of a flight through the Milky Way disc, respecting the distances and apparent luminosity of objects.

\item{\bf GDAS-PR-08}:  I want to produce user friendly interfaces to the Gaia data base which can even be used by students at school or citizen scientists.

\item{\bf GDAS-ST-05}: Standard Queries: filter operations ($==$, $!=$, $<$, $>$, $\le$, $\ge$), combination logic (AND, OR, NOT, XOR); quite often a group of targets to be queried is not aligned along the axes of the parameter space provided by the Archive. Therefore  linear combinations of quantities should also be queriable, e.g. $3 \le a*X + b*Y + c*Z \le 4$ with $a, b, c$ provided by the user and three $X, Y, Z$ quantities archived for every star. 

\item{\bf GDAS-GA-03}: I want whatever measurements there are of Milky Way objects only in a given sky region.

\item{\bf GDAS-BR-15}: I would like to be able to do on-the-fly smoothing/averaging of the data such that queries can be composed giving properties per spatial bin, .e.g.  binned in 1 arcmin elements.

\item{\bf GDAS-SA-01}: I want to get all microlensing events detected in real time by Gaia and see their photometric light curves and astrometric curves.

\item{\bf GDAS-BR-14}: I would like to be able to visualise the Gaia catalogue in multi-dimensional space preferably utilising a visualisation engine such as Google Sky or World Wide Telescope such that I can incorporate data from other surveys into the same tool. 

\item{\bf GDAS-BR-06}: I liked the statistical plots presented in Section 3 of Volume 1 of the Hipparcos and Tycho Catalogues. Show me the same for Gaia and allow me to specify the statistic to explore. 

\item{\bf GDAS-BR-09}: I want to manipulate a 3D cube of the Milky-Way. 

\item{\bf GDAS-GA-14}: I would like to analyse the 5-D, 6-D phase space structure of stellar populations selected by magnitude, colour, abundances in a galactocentric (cylindrical) coordinate system. 

\item{\bf GDAS-PR-01}: For my named constellation, fly me along the line of sight giving me information of interest on each star as I pass it. 

\item{\bf GDAS-ST-24}: Find binary stars with specific colours  where the colours of one or both of the binaries can be specified.

\item{\bf GDAS-ST-07}: Fuzzy Queries. 
Most quantities in Gaia archive will have errorbars. This is relevant for queries, e.g. asking for all stars with $T_{eff}\le 4000$K would not include a star with $T_{eff} = 4050 \pm 250$K. Fuzzy querying should allow this. Required for each quantity: the value, a standard deviation (errorbar), and a probability distribution for this quantity. Most often this will be a Gaussian distribution, but this may not always be the case, particularly if a quantity was computed by dividing two Gaussian quantities. A fuzzy query could add a qualifier to the request: almost certain (P $> 90\%$), likely ($70\% \le$ P $\le 90\%$), perhaps ($30\% \le$ P $ < 70\%$), unlikely ($10\% \le$ P $ < 30\%$), almost certainly not (P $ < 10\%$). The probabilities are computed using the density; the exact definition of the qualifiers could be configured by the user. 

\item{\bf GDAS-GA-17}:  I would like to compare the spatial distribution and dynamics of open star clusters with numerical simulations. 

\item{\bf GDAS-OA-18}: I would like to use standard analysis and visualisation tools Aladin, Topcat, IDL, Python and compare Gaia data with data available in VO standard. 

\item{\bf GDAS-OA-05}: If probabilistic (Bayesian) methods were used to arrive at a particular parameter value please provide the full probability density function and not just value+error. 

\item{\bf GDAS-ED-01}: For my list of stars based on some ground-based observations or catalogue, give me SM CCD image thumbnails. 

\item{\bf GDAS-SA-03}: Fermi has detected a flaring blazar. It has a certain error ellipse, e.g. a few arc-minutes. An optical counterpart is not known. How can I get lightcurves for all objects in the error-ellipse to look for variability and thus possible counterparts to the blazar? 

\item{\bf GDAS-GA-12}:  I want a face-on view of the velocity field(s) of any Galactic objects. 

\item{\bf GDAS-OA-01}: I have a sophisticated simulation code that produces full phase-space realisations of the stellar content of some fraction of the Milky Way. I want to vary certain inputs to my model that are of astrophysical interest and compare against the full Gaia catalogue. Provide me with the means to do this.

\item{\bf GDAS-GA-20}: (same as GDAS-BR-11) I would like to make selections of Gaia objects based not only on Gaia data, but also based on other major catalogues available at that time, such as Pan-STARRS and SDSS. 

\item{\bf GDAS-BR-05}: I want a pretty colour picture for my power-point presentation, created using Gaia data. For example, I would like to select all stars belonging to the halo of the Milky Way and colour them according to their 6D phase space information, assigning to each group its own colour. Then I want to project this on the sky in ecliptic coordinates or any other I choose so that I can illustrate how the halo is structured and where the streams are, or to make a picture like the one from the SDSS, but with Gaia data. I would also like to be able to put labels with the stream names and then save that picture in one of the most common formats without having to actually download the data on my computer. 

\item{\bf GDAS-ST-03}: I want to find clusters of stars in the following way: I select a core set of the stars in the catalogue. The archive then finds me stars that are 'similar' in distance, proper motion, radial velocity, etc. For all these parameters I can set cut-off values that define what 'similar' means. 

\item{\bf GDAS-OA-08}: My target has been observed as part of a large spectroscopic survey (e.g.~from 4MOST and/or WEAVE)  and I would like to be able to interface (in a seamless manner) that external spectroscopic data and the Gaia data.

\item{\bf GDAS-OA-12}: I would like to access the individual astrometric measurements or their residuals with respect to the standard astrometric model similar to the information provided by the  Hipparcos Intermediate Data.

\item{\bf GDAS-OA-17}: I would like to examine the properties of Gaia data in detail without querying the whole Gaia database. 

\item{\bf GDAS-EG-03}: I want the reconstructed image of my favourite source(s). 

\item{\bf GDAS-ST-08}: Pattern Queries: example 'return stars like this set of stars, but avoid stars like that set of stars', where the two sets are defined by the user. Rather challenging to implement, but incredibly useful.

\item{\bf GDAS-ST-10}: Random Queries: return an unbiased random subset of size N of the Gaia Archive, according to a user-specified multi-variate distribution for quantities X, Y, and Z. Computing a reliable histogram, for example, can be done without using the entire archive, as long as you have a unbiased subset of the archive. 

\item{\bf GDAS-ST-11}: Whole-database queries; Sequentially; According to a space-filling curve.

\item{\bf GDAS-ST-12}: Queries in the frequency domain: e.g. return stars with an excess in the Fourier spectrum in the frequency range [x,y]; e.g. return stars with a 1/f noise profile in the Fourier domain.

\item{\bf GDAS-ST-13}: Queries in the time domain: e.g. return stars with at least N points that are 5-$\sigma$ below the median of the light curve. 

\item{\bf GDAS-ST-17}: I have my own private catalogue (of magnitudes, positions, chemical abundances, or other properties) of a globular cluster. I would like to match it with the Gaia catalogue and visualise any of the chosen properties of common objects in a graph, maybe even in 3D form. I also want to do some statistical tests. Maybe in the first exploratory phases I do not want my private catalogue to become public, but I would still like to avoid downloading locally a large chunk of data because I want an instant check of some idea that came when analysing my data. An example: I want a 3D plot of all the stars spatial positions (from Gaia), and I want the stars coloured differently according to their carbon abundance (which I measured). Then I want to project this in different ways and planes, and perform some Kolmogorov-Smirnov statistics, for example.

\item{\bf GDAS-ST-02}: I want to fit a synthetic spectrum to the observed fluxes (BP/RP and/or RVS), as well as additional fluxes from other catalogues (e.g. 2MASS). I would like those data to be directly available in the Gaia catalogue, I don't want to copy/paste from the 2MASS catalogue. I need to be able to degrade the synthetic spectrum with all the Gaia instrumental effects for a useful comparison between observation and theory. 

\item{\bf GDAS-ST-09}: Group Queries: rather than constraining the properties of targets, it is sometimes necessary to constrain the relation between two or more targets, e.g. return all stars with an apparent distance on the sky <= alpha; E.g.: return all stars within this circle on the sky that have the same apex with a tolerance of alpha. 

\item{\bf GDAS-GA-18}: I would like to measure the properties of spiral arms in the extended solar neighbourhood.
\end{itemize}


\section{Appendix B -- Adopted libraries and frameworks}

This appendix includes a list of the most important external libraries and frameworks that were used in the development of, and that integrate,  the Gaia Archive Visualisation Service. 

\begin{table*}[ht]
\caption{External libraries used in Server (indicated with S) and Client (indicated with C).}
\label{tab:libs}
\begin{tabularx}{0.95\textwidth}{XX}
\\\hline
Library & Role \\\hline\noalign{\smallskip}
JUnit \newline \small{\url{http://junit.org/junit4/}} & Java unit test (S)\\    
	\noalign{\smallskip}

Apache Commons I/O \newline \small{\url{https://commons.apache.org/proper/commons-io/}} & I/O helper functions (S)\\        \noalign{\smallskip}

Apache HTTP Client \newline \small{\url{https://hc.apache.org/httpcomponents-client-ga/}} & HTTP client to fetch data from other sites (S)\\    
	\noalign{\smallskip}

Eclipse Jetty \newline \small{\url{http://www.eclipse.org/jetty/}} & Servlet implementation (S)\\    
	\noalign{\smallskip}

EhCache \newline \small{\url{http://www.ehcache.org/}} & Memory smart object cache (S)\\    
	\noalign{\smallskip}

Java JSON \newline \small{\url{http://json.org/}} & JSON  to string conversion and vice versa (S)\\    
	\noalign{\smallskip}

CDS/ZAH ADQL \newline \small{\url{http://cdsportal.u-strasbg.fr/adqltuto/}} & ADQL syntax validation (S)\\    
	\noalign{\smallskip}

STIL \newline \small{\url{httphttp://www.star.bris.ac.uk/~mbt/stil/}} & Dealing with VOTables and FITS tables (S).\\    
	\noalign{\smallskip}

GaiaTools \newline & Dealing GBIN files - Gaia's BINary data format (S).\\    
	\noalign{\smallskip}

angular.js \newline \small{\url{https://angularjs.org/}} & Framework for organising  javascript code in a MVC design pattern (C)\\    
    \noalign{\smallskip}

gridster.js \newline \small{\url{http://dsmorse.github.io/gridster.js/}} & Framework to implement a window-based system (C)\\    
    \noalign{\smallskip}

leaflet.js \newline \small{\url{http://leafletjs.com/}} & Framework to implement a map system (C)\\
    \noalign{\smallskip}

aladin \newline \small{\url{http://aladin.u-strasbg.fr/AladinLite/doc/}} & Aladin lite plug-in (C)\\    
    \noalign{\smallskip}

astro.js \newline \small{\url{http://slowe.github.io/astro.js/}} & Helper functions for astronomic calculations (C)\\    
    \noalign{\smallskip}

bootstrap \newline \small{\url{http://getbootstrap.com/}} & Styling the page (C)\\
    \noalign{\smallskip}

d3.js \newline \small{\url{https://d3js.org/}} & Manipulates documents based on data (e.g. creation of axes, histograms, etc.) (C)\\
    \noalign{\smallskip}

jquery \newline \small{\url{https://jquery.com/}} & Manipulates the HTML of the page (C)\\    
    \noalign{\smallskip}

js9 \newline \small{\url{http://js9.si.edu/}} & JS9 plug-in (C)\\    
    \noalign{\smallskip}

spectrum \newline \small{\url{https://bgrins.github.io/spectrum/}} & Colourpicker manipulation (C)\\    
    \noalign{\smallskip}\hline    \noalign{\smallskip}

\end{tabularx}
\end{table*}

\end{document}